%% file: paper.tex
\documentclass[12pt]{elsarticle}
\usepackage[margin=2.5cm,a4paper]{geometry}
\usepackage{amssymb}
\usepackage{latexsym}
\usepackage{amsmath}
\usepackage{graphics}
\usepackage{nomencl}
\usepackage{color}

\input{format.tex}

\renewcommand{\S}{S_2}
\newcommand{\cF}{\chi}
\newcommand{\prob}[1]{\mathbb{P}\bigl( #1 \bigr)}
\newcommand{\x}{\vek{x}}
\newcommand{\y}{\vek{y}}
\newcommand{\dmn}{\Omega}
\newcommand{\puc}{{\dmn_\Box}}
\newcommand{\ft}[1]{\mathcal{F}\left(#1\right)}
\newcommand{\ift}[1]{\mathcal{F}^{-1}\left( #1 \right)}
\newcommand{\p}{\vek{p}}
\renewcommand{\k}{\vek{k}}
\newcommand{\tmp}{\theta}
\newcommand{\Tmp}{\Theta}
\newcommand{\Tmpgrad}{\vek{\nabla}\Tmp}
\newcommand{\tmpgrad}{\vek{\nabla}\tmp}
\newcommand{\q}{\vek{q}}
\newcommand{\Q}{\vek{Q}}
\newcommand{\ctc}{\vek{\lambda}}

\newcommand{\gfun}{\vek{\Gamma}}
\renewcommand{\hom}{^\mathrm{hom}}

\makenomenclature

\journal{arxiv}

\begin{document}

%%%%%%%%%%%%%%%%%%%%%%%%%%%%%%%%%%%%%%%%%%%%%%%%%%%%%%%%%%%%%%%%%
\begin{frontmatter}
\title{Selected topics in homogenization of transport processes in historical masonry structures}

\author[ctu]{Jan S\'{y}kora}
\ead{jan.sykora.1@fsv.cvut.cz}

\author[ctu,it4]{Jan Zeman}
\ead{zemanj@cml.fsv.cvut.cz}

\author[ctu]{Michal \v{S}ejnoha\corref{auth}}
\ead{sejnom@fsv.cvut.cz}
\cortext[auth]{Corresponding author. Tel.:~+420-2-2435-4494;
fax~+420-2-2431-0775}
\address[ctu]{Department of Mechanics, Faculty of Civil Engineering,
  Czech Technical University in Prague, Th\'{a}kurova 7, 166 29 Prague
  6, Czech Republic}
\address[it4]{Centre of Excellence IT4Innovations, V\v{S}B-TU Ostrava, 
17.~listopadu 15/2172 708 33 Ostrava-Poruba, Czech Republic}

\begin{abstract}
{ The paper reviews several topics associated with the homogenization
  of transport processed in historical masonry structures. Since these
  often experience an irregular or random pattern, we open the subject
  by summarizing essential steps in the formulation of a suitable
  computational model in the form of Statistically Equivalent Periodic
  Unit Cell (SEPUC).  Accepting SEPUC as a reliable representative
  volume element is supported by application of the Fast Fourier
  Transform to both the SEPUC and large binary sample of real masonry
  in search for effective thermal conductivities limited here to a
  steady state heat conduction problem. Fully coupled non-stationary
  heat and moisture transport is addressed next in the framework of
  two-scale first-order homogenization approach with emphases on the
  application of boundary and initial conditions on the meso-scale. }
\end{abstract}

\end{frontmatter}

%%%%%%%%%%%%%%%%%%%%%%%%%%%%%%%%%%%%%%%%%%%%%%%%%%%%%%%%%%%%%%%%%%%%%%%%%%%%%
\section{Introduction}
%%%%%%%%%%%%%%%%%%%%%%

Advanced computational simulations of historic structures are becoming
increasingly common in the assessment of their existing state and in
planning of reconstruction measures~\cite{Lourenco:2002:CHMS}. In this
context, particular attention needs to be paid to variations of
temperature and moisture fields, whose contribution to structural
damage usually far exceeds the effects of mechanical loadings, see
e.g.~\cite{Rankin:2005:TEC,witzany:2008:FRH,Fajman:2010:ITC} for
concrete case studies. Taking into account highly heterogeneous
character of historical constructions, simulations of these phenomena
often necessitates the deployment of multi-scale strategies, developed
for
mechanical~\cite{Anthoine:IJSS:95,Lourenco:2007:AMS,Massart:2007:EMS,Sejnoha:2008:MSHM}
and transport processes~\cite{Sykora:2008:ACHM,Sykora:2011:AMC} in
masonry structures. Successful engineering application of this
modeling approach to the assessment of Charles bridge in Prague is
described in our work~\cite{Zeman:2008:PMSMP}. There,
three-dimensional multi-physics analysis of the bridge body was
executed, with material parameters at the macroscopic structural scale
determined from meso-scale simulations. Among other things, results of
the study highlighted the need for fully coupled macro-meso
simulations of transport processes. In this paper, we complement these
results by a more detailed analysis of two aspects of multi-scale
simulations, namely the \emph{meso-scale representation} for irregular
masonry structures and the introduction of proper \emph{boundary and
  initial conditions} in the macro-meso scale transition.

The masonry texture at meso-scale is to be incorporated in the form of
the Representative Volume Element~(RVE), a statistically
representative sample of the analyzed
material~\cite{Hill:1963:EPT}. Specification of the RVE is
particularly simple in case of a regular masonry, for which it reduces
to a Periodic Unit Cell~(PUC) associated to a given type of the
bond~\cite{Anthoine:IJSS:95}. However, historical masonry structures
are typical of irregular or random textures, which renders the
determination of PUC unambiguous.  A convenient approach to overcome
this difficulty was introduced by Povirk~\cite{Povirk:1995:IMI}, who
suggested to replace the original complex meso-structure with an
idealized PUC, with parameters determined by matching spatial
statistics of original and simplified representation. By combining his
ideas with related works on microstructure reconstruction,
e.g.~\cite{Yeong:1998:RRM}, the concept of Statistically Equivalent
Periodic Unit cell~(SEPUC) was later successfully applied to,
e.g. analysis of
fiber-reinforced~\cite{Zeman:2001:EPG,Sejnoha:2002:OVR} and woven
composites~\cite{Zeman:2004:HBWCI,Vorel:2012:HPW}, high-density
poly-disperse particle packings~\cite{Lee:2009:TDR} or
micro-heterogeneous steels~\cite{Schroder:2011:ARM}. In
\secref{sec:geo_model}, we employ this procedure to an irregular
masonry wall of Charles Bridge in Prague, characterized by a digital
photograph.

Generating a suitable RVE of a heterogeneous material is just the
first step towards a reliable prediction of material as well as
structural response of masonry. Focusing on the description of
transport processes in heterogeneous media the reader is
advised to study the work by \"{O}zdemir et
al.~\cite{Ozdemir:IJNME:2008} and Larsson et
al.~\cite{Larsson:2010:IJMNE}. Through the application of consistent
variational formulation the authors in~\cite{Larsson:2010:IJMNE}
suggested the macroscopic response to be dependent on the actual size of
mesoscopic RVE, providing the transient conditions are assumed on both
the macro- and meso-scale. Being aware of the need for a fully coupled
multi-scale analysis of simultaneous heat and moisture transport in
masonry structures performed recently in~\cite{Sykora:JCAM:2011}, it
becomes clear that additional sub-stepping of a given macroscopic time
step on the meso-scale may considerably slow down the computational
process. To show that in some cases this step might be
avoided by running the meso-scale analysis under steady state
conditions~\cite{Ozdemir:IJNME:2008} even for a finite size RVE may
thus prove useful in keeping the computational cost relatively
low. This issue is addressed in the second part of this paper,
Section~\ref{sec:coupled}, with particular attention dedicated to the
influence of loading and initial conditions imposed on meso-scale.

\section{Image-based geometrical modeling}\label{sec:geo_model}
%%%%%%%%%%%%%%%%%%%%%%%%%%%%%%%%%%%%%%%%%% 
This section deals with the first aspect of multi-scale simulations
announced above, namely with the realistic representation of irregular
masonry structure relying on image-based data. To this purpose, in
\secref{sec:SEPUC_def} we introduce the model of an idealized
mesostructure and the procedure to determine its parameters.
\secref{sec:fft_hom} briefly reviews the numerical scheme employed to
determine local fields operating directly on mesostructural
images. These two tools are combined together in
\secref{sec:SEPUC_example} to assess the SEPUC quality in view of the
distribution of heat fluxes in disordered masonry under steady-state
conditions.

\subsection{Strategy of SEPUC determination}\label{sec:SEPUC_def}
%%%%%%%%%%%%%%%%%%%%%%%%%%%%%%%%%%%%%%%%%%%%
The key step in the SEPUC definition is a proper choice of the spatial
statistics to characterize the dominant features of the heterogeneous material
under study. With the focus on masonry structures,  we limit our attention to
the \emph{mortar} two-point probability function, see
e.g.~\cite{Torquato:2002:RHM} for a general overview and examples of alternative
statistical descriptors.

To introduce the subject, consider a masonry sample $\dmn$ composed of mortar
and blocks, and denote the characteristic function of the domain occupied by mortar
as $\cF(\x)$. Then, the two-point probability function $\S$ states the
probability that two points $\x$ and $\y$, randomly thrown into a medium,
will both be found in the mortar phase:
\nomenclature{$\y$}{Point used to expressed $\S$}
\begin{equation}
\S( \x, \y ) 
=
\prob{\cF(\x) \cF(\y) = 1}.
\end{equation}
\nomenclature{$\cF$}{Characteristic function}%
\nomenclature{$\S$}{Two-point probability}%
\nomenclature{$\prob{x}$}{Probability of event $x$}%
For the case of statistically homogeneous and ergodic media, two-point
probability function depends on $(\x - \y)$ only and $\S(\vek{0}) = \volfrac$,
where $\volfrac$ is the mortar volume fraction. Moreover, it can be obtained
from the relation
\nomenclature{$\volfrac$}{Volume fraction}%
\begin{equation}\label{eq:S2_FFT}
\S
=
\frac{1}{\measure{\dmn}}
\ift{%
\ft{\cF}
\overline{\ft{\cF}}
},
\end{equation}
that can be efficiently evaluated using the Fast Fourier Transform~(FFT)
techniques for image-based
microstructures~\cite{Gajdosik:2006:QAFC,Torquato:2002:RHM}. In
\Eref{eq:S2_FFT}, $\ft{\bullet}$ and $\ift{\bullet}$ designate the forward and
inverse Fourier Transform operators, respectively, $\overline{\bullet}$ denotes
the complex conjugate and $\measure{\dmn}$ is the area of $\dmn$.
\nomenclature{$\puc$}{Analyzed domain}%
\nomenclature{$\ft{\bullet}$}{Forward Fourier transform}%
\nomenclature{$\ift{\bullet}$}{Inverse Fourier transform}%
\nomenclature{$\measure{x}$}{Measure of $x$}%
\nomenclature{$\overline{x}$}{Complex conjugate of $x$}%

Once the original structure has been quantified by a suitable
statistical descriptor, a proper parametrization of the idealized cell
geometry needs to be introduced, expressed here by a parameter vector
$\p$. Its optimal value then follows from the minimization of least
square error
\begin{equation}\label{sec:SEPUC_objfunc}
E(\p)
=
\frac{1}{\measure{\puc}}
\int_\puc
\left( \overline{\S}(\x) - \S(\x,\p) \right)^2
\de \x,
\end{equation}
expressed as the difference between the target statistical descriptor
$\overline{\S}(\x)$ related to the original microstructure and the unit cell
associated with $\p$, integrated over the unit cell domain $\puc \subset \dmn$.
\nomenclature{$\p$}{Parameter vector}%
\nomenclature{$E$}{Error function}%

A closer inspection reveals that the objective
function~\eqref{sec:SEPUC_objfunc} is non-convex, multimodal and
discontinuous due to the effect of limited bitmap resolution. Based on
our previous experience~\cite{Hrstka:2003:CCD,Matous:2000:GEI}, a
global stochastic optimization algorithm, relying on the combination
of real-valued genetic algorithms and the Simulated Annealing method,
is employed to solve this optimization problem.

\subsection{Homogenization scheme}\label{sec:fft_hom}

The distribution of local fields within $\puc$ follows from the solution of
periodic unit cell problem~\cite{Milton:2002:TC}
\begin{eqnarray}\label{eq:unit_cell_problem}
\vek{\nabla} \times \tmpgrad = \vek{0},
&
\displaystyle
\vek{\nabla} \cdot \q = 0,
&
\q 
=
- \ctc
\tmpgrad
\mbox{ in } \puc,
\end{eqnarray}
in which $\q$ stands for the thermal flux vector, $\ctc$ is the second-order
tensor of material conductivity and $\tmpgrad$ denotes the $\puc$-periodic
field of temperature gradient satisfying
\begin{equation}
\frac{1}{|\puc|}
\int_{\puc}
\tmpgrad( \x )
\de \x
=
\Tmpgrad,
\end{equation}
where $\Tmpgrad$ is the macroscopic temperature gradient prescribed over $\puc$,
see \secref{sec:coupled} below for additional details. It is
well-known that the solution to the unit
cell satisfies the Lippman-Schwinger equation~\cite{Milton:2002:TC}
\begin{equation}\label{eq:int_equation}
\tmpgrad( \x )
+
\int_{\puc}
\gfun
(\x - \y )
\delta \ctc(\y)
\tmpgrad(\y)
\de\y
=
\Tmpgrad,
\end{equation}
where $\delta \ctc = \ctc - \lambda^{(0)} \vek{I}$, $\lambda_0$ is the
conductivity of an auxiliary isotropic reference medium and the second-order
operator $\gfun$ is related to the Green function
of the problem~\eqref{eq:unit_cell_problem} with $\ctc(\x) = \lambda^{(0)}
\vek{I}$. It admits a compact closed-form expression in the Fourier
space~\cite{Milton:2002:TC}
\begin{equation}
\ft{\gfun}( \k )
=
\begin{cases}
\vek{0} & \mbox{for } \k = \vek{0},
\\
\displaystyle
\frac{\k \otimes \k}{\k \cdot \k} & \mbox{otherwise},
\end{cases}
\end{equation}
so that its action can be efficiently evaluated by the FFT algorithm. This
observation resulted in an iterative scheme due to Moulinec and
Suquet~\cite{Moulinec:1994:FNMC}, applicable to arbitrary digitized
media.

In our case, we adopt an accelerated version of the original algorithm
due to Zeman \emph{et al.}~\cite{Zeman:2010:AFFT}. Since the sample is
discretized by a regular $N_1 \times N_2$ bitmap, it is convenient to
project the integral equation~\eqref{eq:int_equation} onto the space
of trigonometric polynomials, e.g.~\cite{Saranen:2002:PIP}. This
yields the linear system in the form
\begin{equation}\label{eq:FFT_system}
( \vek{I} + \vek{B} ) \nabla\vek{\tmp}_{\mathrm{d}} = \nabla\vek{\Tmp}_{\mathrm{d}},
\end{equation}
where the $2 N_1 N_2$ vector $\nabla\vek{\tmp}_{\mathrm{d}}$ stores the unknown discrete values
of temperature gradient at pixels, $\nabla\vek{\Tmp}_{\mathrm{d}}$ is the corresponding vector of
the overall temperature gradient and matrix $\vek{B}$ is expressed as
\begin{equation}
\vek{B}
= 
\begin{pmatrix}
\vek{F}^{-1} & \vek{0} \\
\vek{0} & \vek{F}^{-1} 
\end{pmatrix}
\begin{pmatrix}
\gfun_{11} & 
\gfun_{12} \\ 
\gfun_{21} & 
\gfun_{22}  
\end{pmatrix}
\begin{pmatrix}
\vek{F} & \vek{0} \\
\vek{0} & \vek{F} 
\end{pmatrix}
\begin{pmatrix}
\delta\ctc_{11} & 
\delta\ctc_{12} \\
\delta\ctc_{21} & 
\delta\ctc_{22}  
\end{pmatrix}.
\end{equation}
Here, the matrices $\vek{F}$ and $\vek{F}^{-1}$
implement the forward and the inverse discrete Fourier transform and, e.g.
$\delta\ctc_{12}$ is a diagonal $(N_1 N_2) \times (N_1 N_2)$ matrix 
storing the corresponding component of the conductivity matrix at individual
pixels, see~\cite{Zeman:2010:AFFT} for more details. The system~\eqref{eq:FFT_system} is
solved using standard conjugate gradient algorithm.

Upon convergence, the distribution of the local heat flux $\q$
due to $\Tmpgrad$ is determined from the solution $\nabla\vek{\tmp}_{\mathrm{d}}$ by
\Eref{eq:unit_cell_problem}$_3$ and the global heat flux is computed as
\begin{equation}
\Q 
=
\frac{1}{\measure{\puc}}
\int_{\puc}
\q(\x)
\de \x.
\end{equation}
\nomenclature{$\Q$}{Macroscopic heat flux}%
This allows us to determine the $2\times 2$ homogenized conductivity
matrix $\ctc\hom$ via the solution of two successive steady state heat
conduction problems. To that end, the periodic unit cell is loaded, in
turn, by each of the two components of $\Tmpgrad$ equal to unity,
while the remaining one vanishes. The corresponding volume flux
averages $\Q$ then provide individual columns of~$\ctc\hom$.

\subsection{Example}\label{sec:SEPUC_example}

Principles of the introduced methodology are illustrated by the
analysis of sandstone facing masonry wall of Charles Bridge in Prague
appearing \figref{fig:original_structure}. As the first step, the
original color image, \figref{fig:original_structure}(a), was
thresholded to the binary representation, manually adjusted to remove
image processing artifacts and a rectangular domain $\dmn$ was
selected for further analysis, \figref{fig:original_structure}(b).

\begin{figure}[ht]
\label{fig:original_structure}
\centering
\begin{tabular}{cc}
\includegraphics[height=5.6cm]{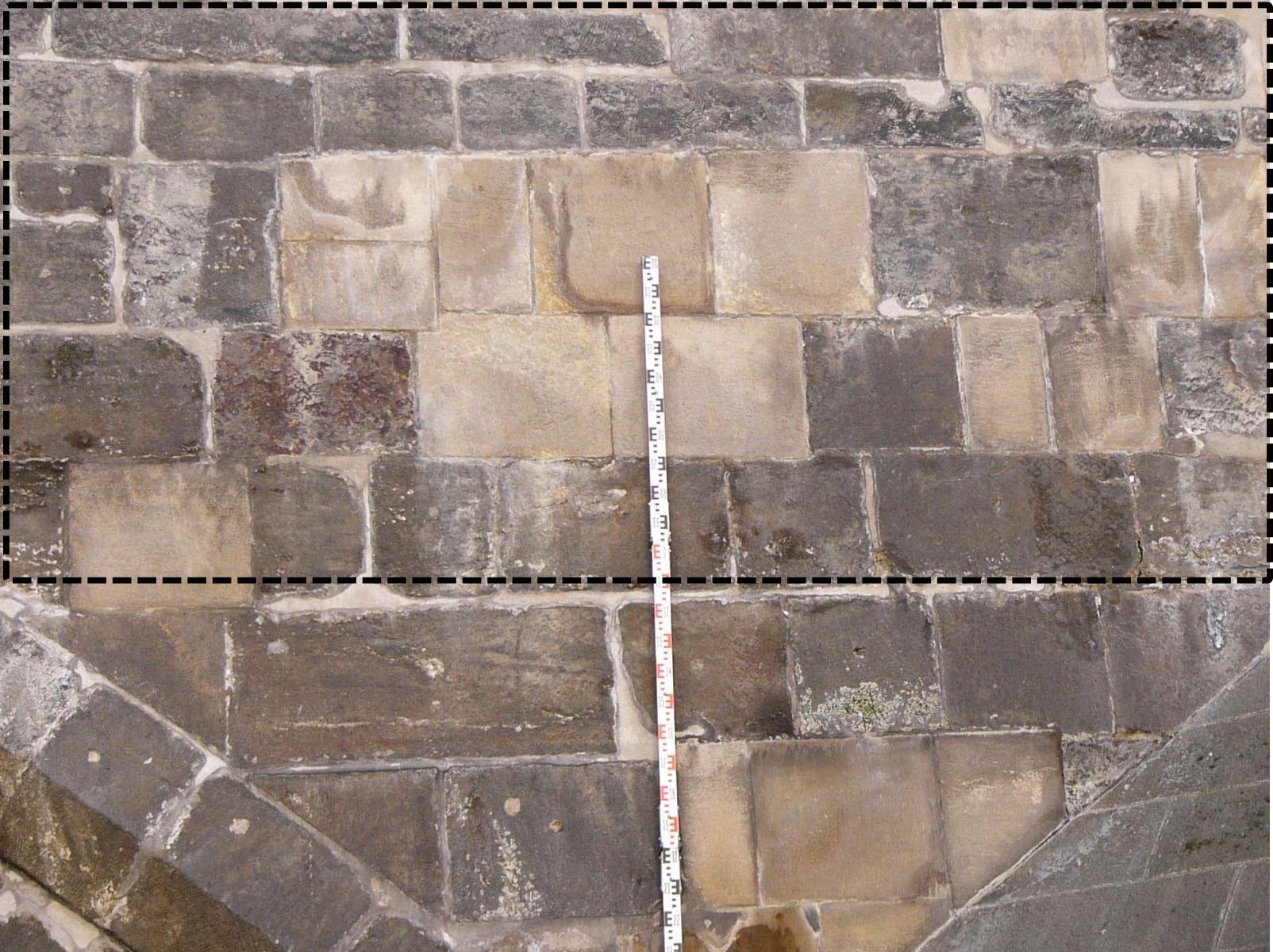} &
\includegraphics[height=5.6cm]{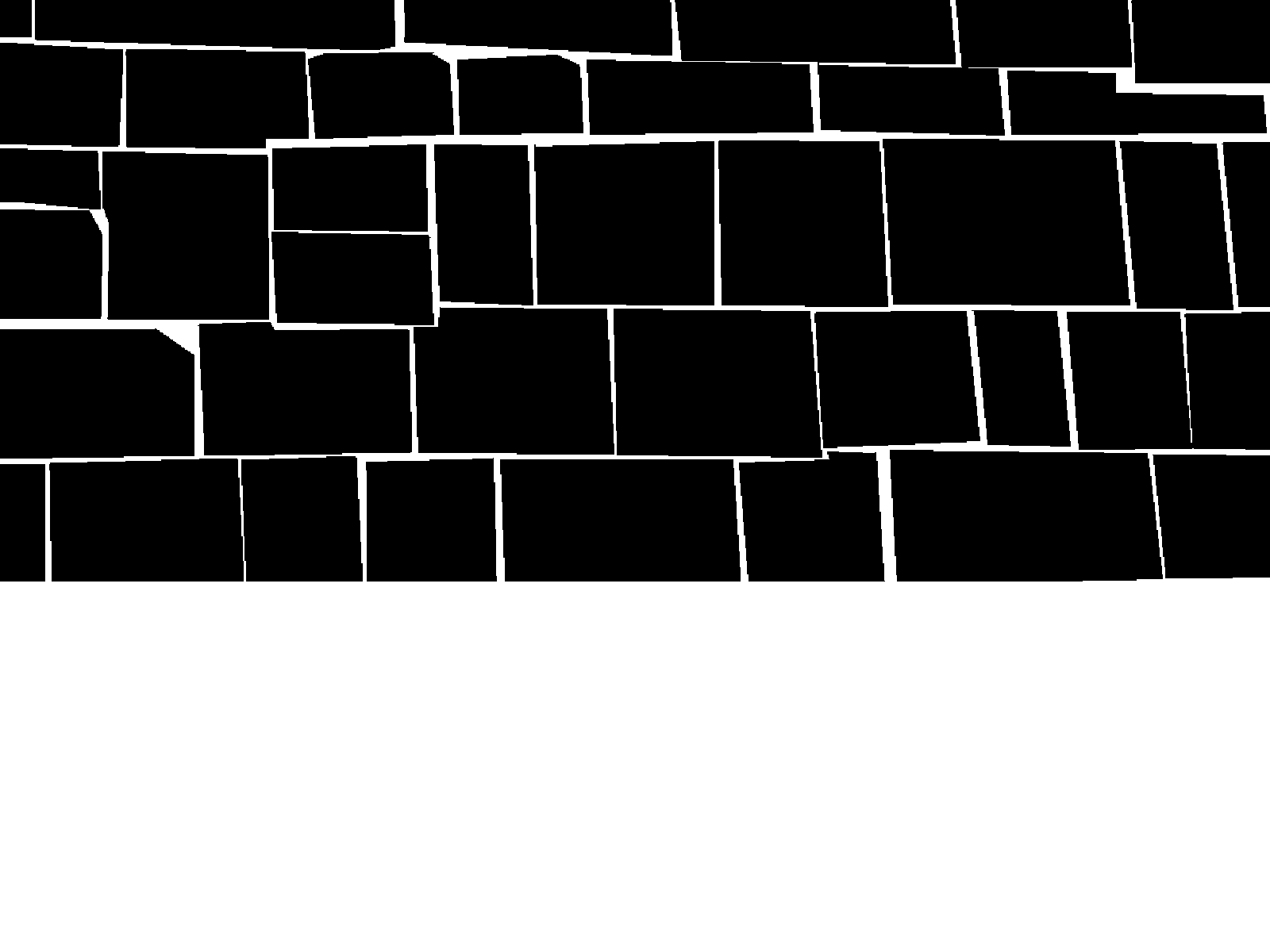} \\
(a) & (b)
\end{tabular}
\caption{(a)~Example of a masonry wall of Charles Bridge in Prague and
(b)~meso-structural window $\dmn$ provided in the form of $1,600
\times 735$ black-and-wide bitmap}
\end{figure}

Our aim is to replace this complex meso-structure with its idealized
representation in terms of a two-block-layer SEPUC, described by twelve
parameters assembled in the vector
\begin{equation}
\p 
=
\begin{pmatrix}
b, b_1, b_2, \Delta, h_1, h_2, t_1, t_2, t_3, t_4, t_5, t_6 
\end{pmatrix},
\end{equation}
see \figref{fig:SEPUC}(a). These parameters are adjusted to minimize the
discrepancy between the target two-point probability function,
\figref{fig:SEPUC_S}(a), and the one corresponding to the SEPUC,
\figref{fig:SEPUC_S}(b). As visible, the resulting SEPUC-based representation
$\puc$, \figref{fig:SEPUC}(b), captures the dominant features of the original
microstructures, such as the overall volume fraction equal to $7.5\%$, or the
average thickness of joints, corresponding to the spread of the peak at $\x =
\vek{0}$ in the horizontal and vertical directions. However, a significant amount
of fine-scale features of the two-point probability functions, visible in the
original data, are filtered out or poorly reproduced by the adopted, more
regular, representation.

\begin{figure}[p]
\centering
\begin{tabular}{cc}
\includegraphics[height=60mm]{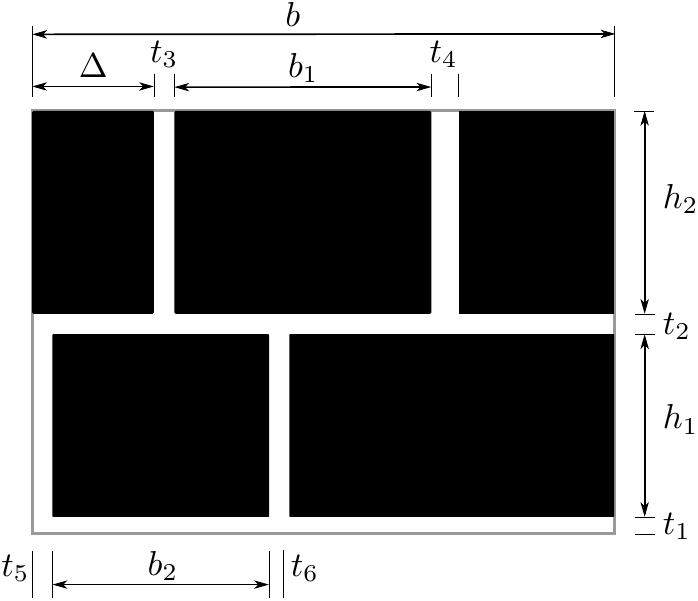} &
\includegraphics[height=60mm]{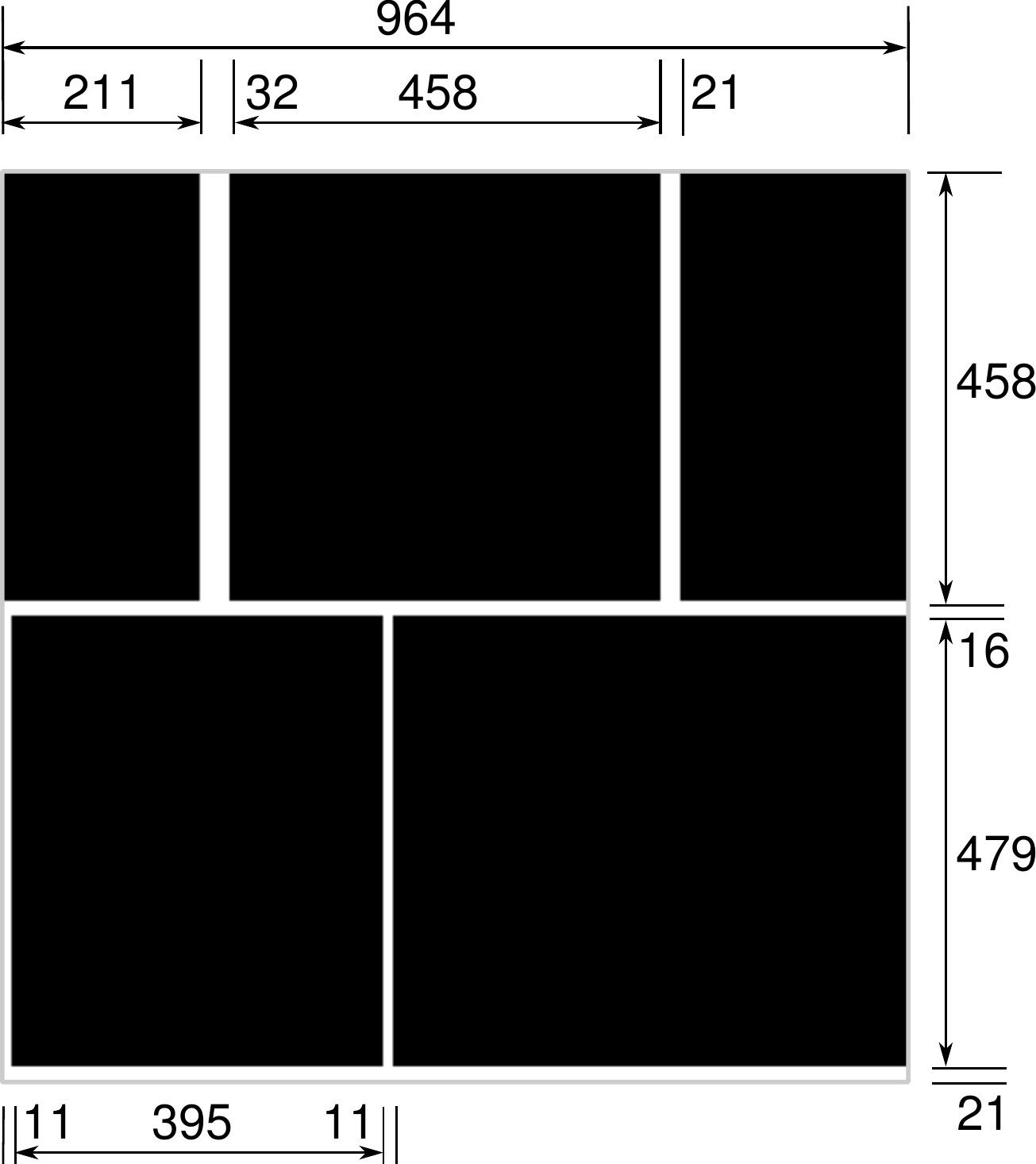}
\\
(a) & (b)
\end{tabular}
\caption{Determination of SEPUC for irregular masonry; (a)~adopted
parametrization and (b)~optimal mesostructure with dimensions in [mm]}
\label{fig:SEPUC}
%\end{figure}
%\begin{figure}[ht]
\centering
\begin{tabular}{cc}
\includegraphics[height=56.5mm]{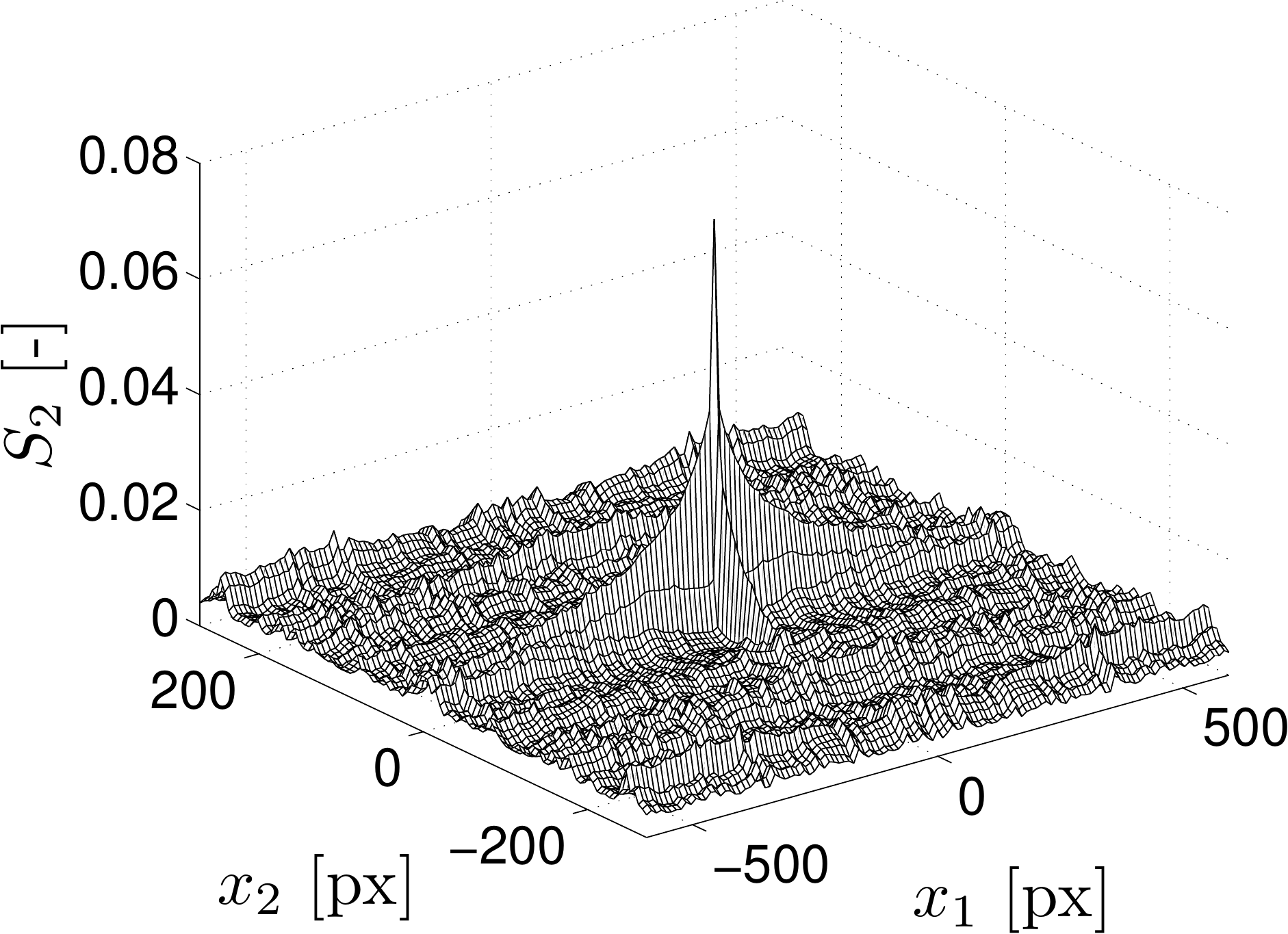} &
\includegraphics[height=56.5mm]{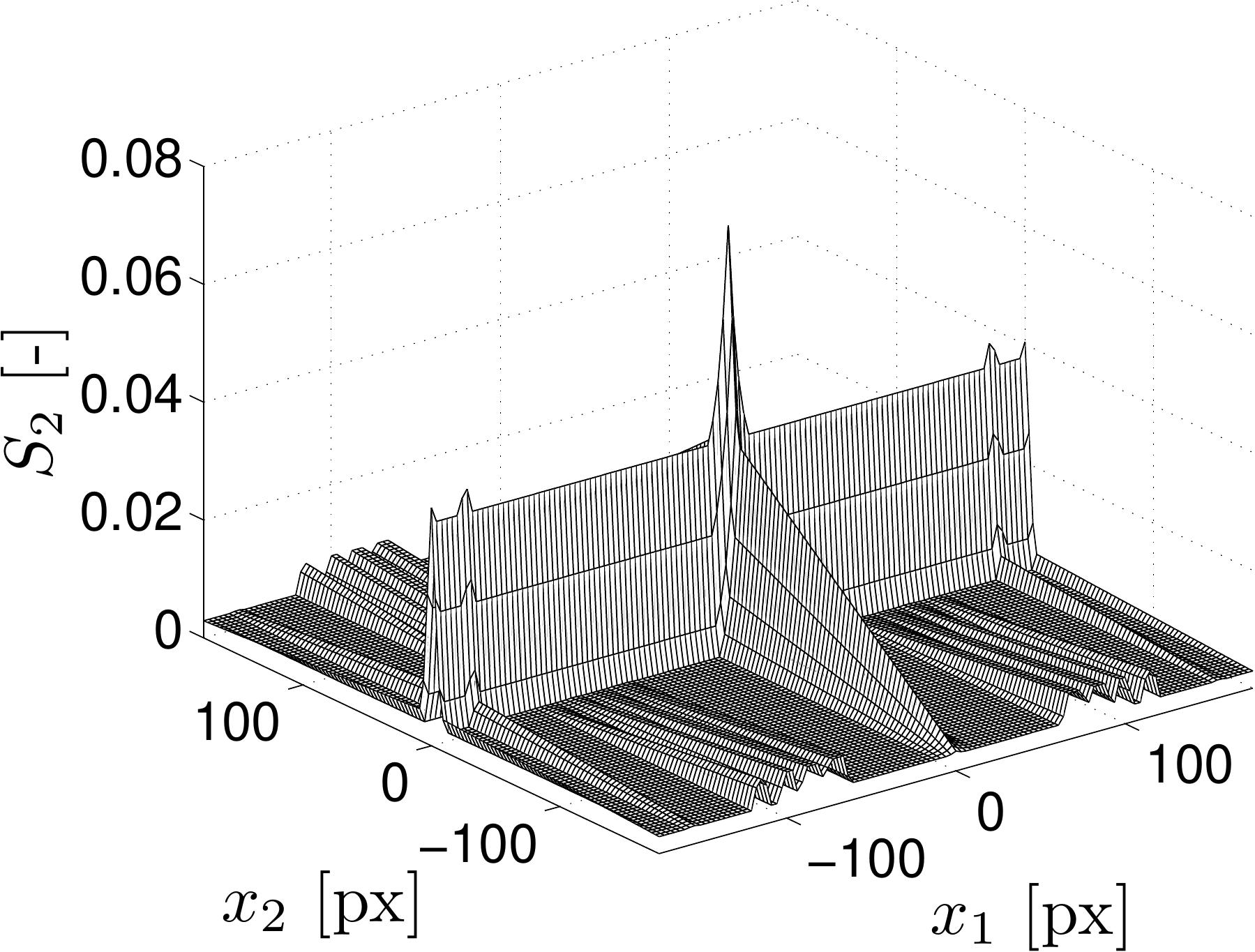} \\
(a) & (b)
\end{tabular}
\caption{Two-point probability function describing (a)~target microstructure and
(b)~SEPUC (1 pixel corresponds to $\sim 2.6$~mm)}
\label{fig:SEPUC_S}
%\end{figure}
%\begin{figure}[ht]
\centering
\begin{tabular}{cc}
\includegraphics[height=52.5mm]{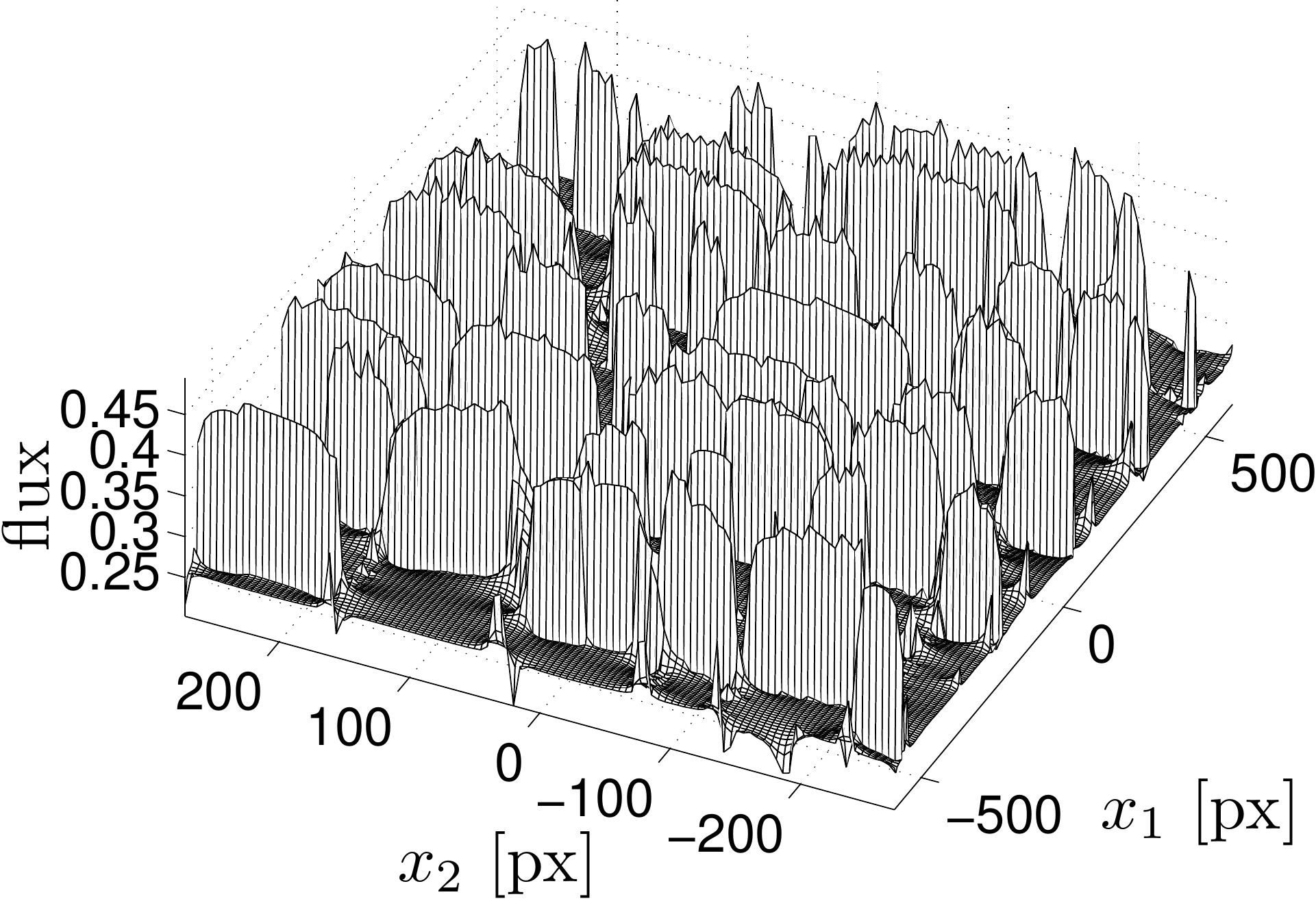} &
\includegraphics[height=52.5mm]{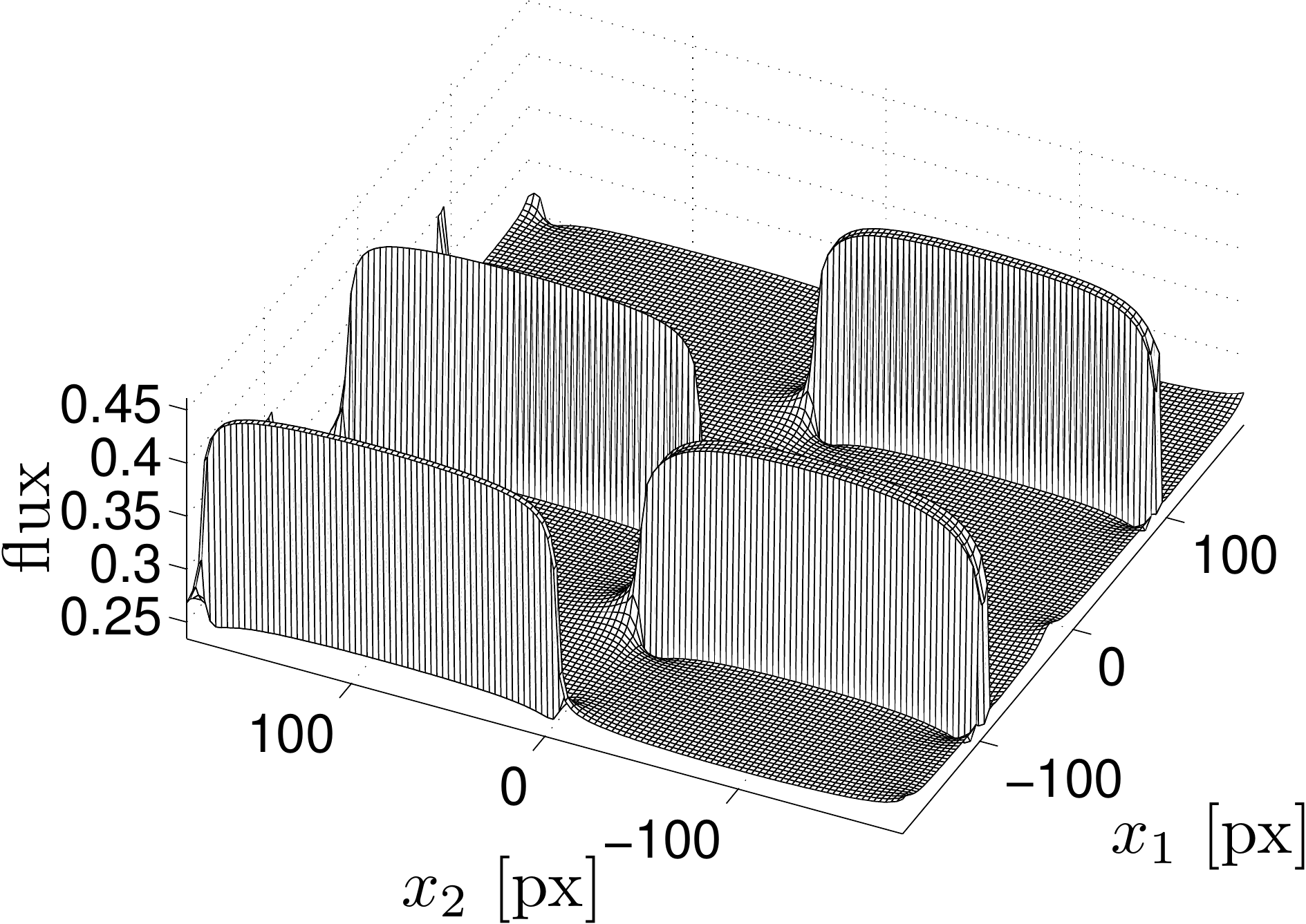} \\
(a) & (b)
\end{tabular}
\caption{Distribution of heat flux magnitudes $\| \q \|$~(in Wm$^{-1}$) due to
due to macroscopic temperature gradient $\partial \Tmp / \partial x_2 =
1$~Km$^{-1}$ in (a)~original domain and~(b) SEPUC}
\label{fig:SEPUC_local_fields}
\end{figure}

Of course, the impact of such approximation needs to be quantified
a-posteriori, from the point of view of the particular application. In
our case, such a step is performed on the basis of homogenized
conductivities and distribution of heat fluxes as determined by the
FFT scheme. The distribution of the magnitudes of local fields due to
macroscopic temperature gradient $\partial \Tmp / \partial x_2$
appears in \figref{fig:SEPUC_local_fields}, for phase conductivities
set according to \tabref{tab:matpar}. We observe that even though the
local fields within the SEPUC are certainly more regular than in the
original mesostructure, the extreme values as well as the average
distributions are reproduced surprisingly well.

This claim is further supported by very close match between the
homogenized conductivities of SEPUC and the original media,
\tabref{tab:homog_conductivites}, demonstrating that the SEPUC is
capable of reproducing almost perfect isotropy of the original
sample. Finally, for the sake of comparison, we also present
homogenized conductivities of unit cells used in the following
section, cf. \figref{fig:scheme}. The influence of the degree of
heterogeneity on the resulting predictions is evident. 

\begin{table}[ht]
\centering	
\begin{tabular}{lrrr}
\hline
& $\lambda\hom_{11}$ & $\lambda\hom_{12} = \lambda\hom_{21}$ &
$\lambda\hom_{22}$
\\
\hline
Original mesostructure & 
0.2612 & 0.0000 & 0.2622 \\
SEPUC & 
0.2616 & 0.0000 & 0.2618 \\
\hline
Regular masonry &
0.2856 & 0.0000 & 0.2967 \\ 
Irregular masonry & 
0.2932 & -0.0015 & 0.2960 \\
\hline
\end{tabular}
\caption{Homogenized thermal conductivities in [$\unitConduct$]}
\label{tab:homog_conductivites}
\end{table}

\section{Multi-scale homogenization of coupled heat and moisture transport}\label{sec:coupled}
%%%%%%%%%%%%%%%%%%%%%%%%%%%%%%%%%%%%%%%%%%%%%%%%%%%%%%%%%%%%%%%%%%%%%%%%%%%%%%%%%%%%%%%%%%%%%%%%%%%%%%%%
In this section, we continue with the selected example of Charles
Bridge and extend the previous study to the multi-scale modeling of
coupled nonlinear transient heat and moisture transport in masonry
structures. While still adopting the first order homogenization
approach (linear variation of macroscopic temperature and moisture
fields is assumed) we choose, unlike the previous section, the finite
element method (FEM) to solve the resulting system of partial
differential equations. In the present study, these arise from the
application of a nonlinear diffusion model proposed by K\"{u}nzel
in~\cite{Kunzel:1995,Kunzel:IJHMT:1997}.  The model is described by
the energy balance equation
\begin{equation}
\frac{\mathrm{d}H}{\mathrm{d}\theta}\frac{\mathrm{d}\theta}{\mathrm{d}t}
 =  \vek{\nabla}^{\mathrm{T}}[\lambda\vek{\nabla} \theta]+h_{v}\vek{\nabla}^{\mathrm{T}}
[\delta_{p}\vek{\nabla}\{\varphi p_{\mathrm{sat}}(\theta)\}] \, ,
\label{eq:hom01}
\end{equation}
and by the mass conservation equation
\begin{equation}
\frac{\mathrm{d}w}{\mathrm{d}\varphi}\frac{\mathrm{d}
\varphi}{\mathrm{d}t}  =  \vek{\nabla}^{\mathrm{T}}
[D_{\varphi}\vek{\nabla}\varphi]+\vek{\nabla}^{\mathrm{T}}
[\delta_{p}\vek{\nabla}\{\varphi p_{\mathrm{sat}}(\theta)\}] \, ,
\label{eq:hom01m}
\end{equation}
where temperature $\theta$ and relative humidity $\varphi$ are the two
state variables. The model requires specifying the enthalpy of the
moist building material $H$, the water content of the building
material $w$, the coefficient of thermal conductivity $\lambda$, the
liquid conduction coefficient $D_{\varphi}$, the water vapor
permeability $\delta_{p}$, the evaporation enthalpy of water $h_{v}$
and the water vapor saturation pressure $p_{\mathrm{sat}}$.  Details
regarding the functional dependence of the model parameters on
temperature and relative humidity are available, e.g. in
~\cite{Kunzel:1995,Sykora:2010}.

%%%%%%%%%%%%%%%%%%%%%%%%%%%%%%%%%%%%%%%%%%%%%%%%%%%%%%%%%%%%%%%%%
\begin{figure} [ht!]
\begin{center}
\begin{tabular}{c@{\hspace{5mm}}}
\includegraphics*[width=95mm,keepaspectratio]{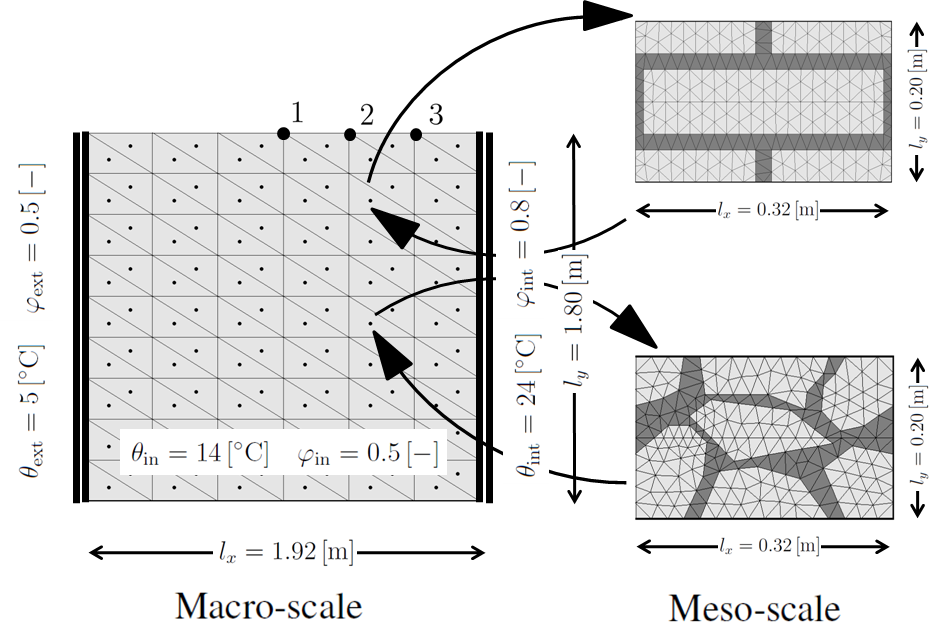}
\end{tabular}
\end{center}
\caption{Scheme of coupled multi-scale framework}
\label{fig:scheme}
\end{figure}
%%%%%%%%%%%%%%%%%%%%%%%%%%%%%%%%%%%%%%%%%%%%%%%%%%%%%%%%%%%%%%%%%

In~\cite{Sykora:2010} the authors noticed a significant dependence of
the homogenized macroscopic properties on the applied temperature and
relative humidity gradients. This finding promoted the solution of
full scale analysis of masonry structures in a fully coupled
multi-scale homogenization framework. To that end, a nested FE$^2$
scheme, graphically presented in Figure~\ref{fig:scheme} (see
also~\cite{Ozdemir:IJNME:2008,Larsson:2010:IJMNE,Sykora:JCAM:2011} for
more details), appears as a suitable method of attack. In such a case
we expect the homogenized macro-scale fields to be found from the
solution of a certain sub-scale (meso-scale) problem performed on an
RVE loaded by the prescribed constant temperatures and moisture
gradients.

As already mentioned in the introductory part, Larson et
al.~\cite{Larsson:2010:IJMNE} have shown that assuming a
non-stationary response also on meso-scale generates a non-local term
in the homogenized macroscopic equations, which renders the
macroscopic response dependent on the RVE size. There is no dispute
that performing the non-stationary analysis on both scales may
considerably increase the computational cost making the numerical
analysis prohibitively expensive even if exploiting
parallelization. To reconcile these issues for a typical masonry
material and RVE sizes is thus crucial for the success of analysis of
full scale tree-dimensional models of historical masonry structure
such as Charles Bridge. This will be the main topic of subsequent
paragraphs addressing also the effect of loading, boundary and initial
conditions.

Since irrelevant from the computational point of view we consider,
henceforth, only the regular (PUC) and irregular (SEPUC) bonding of
masonry displayed in Figure~\ref{fig:scheme}.

\subsection{Theoretical formulation}
%%%%%%%%%%%%%%%%%%%%%%%%%%%%%%%%%%%%%%%%%%%%%%%%%%%%%%%%%%%%%%%%%%%%%%%%%%%%%%
To begin, we adopt a variationally consistent homogenization outlined
in detail in~\cite{Larsson:2010:IJMNE} starting with the assumption
that a local field $a$ can be replaced by a spatially homogenized one
$\langle{a}\rangle$ such that
\begin{eqnarray}
\int_{\Omega}a\,\mathrm{d}\Omega & \approx &
\int_{\Omega}\left\langle a\right\rangle_{\Box}
\mathrm{d}\Omega=\int_{\Omega}\left(\frac{1}{\left|\Omega_{\Box}\right|}\int_{\Omega_{\Box}}
a\,\mathrm{d}\Omega_{\Box}\right)\mathrm{d}\Omega, \label{eq:hom02} \\
\int_{\Gamma}a\,\mathrm{d}\Gamma & \approx &
\int_{\Gamma}\left\langle a\right\rangle_{\Box}
\mathrm{d}\Gamma=\int_{\Gamma}\left(\frac{1}{\left|\Gamma_{\Box}\right|}\int_{\Gamma_{\Box}}
a\,\mathrm{d}\Gamma_{\Box}\right)\mathrm{d}\Gamma,
\label{eq:hom03}
\end{eqnarray}
where $\Omega_{\Box}$ and $\Gamma_{\Box}$ represent the internal and
boundary parts of PUC. In what follows, for the sake of
lucidity, we shall treat only the energy balance
equation~\eqref{eq:hom01} which upon employing Eqs.  (\ref{eq:hom02})
and (\ref{eq:hom03}) becomes
\begin{eqnarray}
&&\int_{\Omega}\left\langle\delta
\theta\frac{\mathrm{d}H}{\mathrm{d}\theta}\frac{\mathrm{d}\theta}{\mathrm{d}t}\right\rangle_{\Box}\mathrm{d}\Omega
+\int_{\Omega}\left\langle\{\vek{\nabla}\delta
\theta\}^{\mathrm{\textsf{T}}}\left[\{\mat{\lambda}\vek{\nabla}
\theta\}+h_{v}
\left\{\mat{\delta}_{p}\varphi\frac{\mathrm{d}p_{\mathrm{sat}}}{\mathrm{d}\theta}\vek\nabla\theta\right\}\right]\right\rangle_{\Box}\mathrm{d}\Omega +\nonumber \\
&& +\int_{\Omega}\left\langle\{\vek{\nabla}\delta
\theta\}^{\mathrm{\textsf{T}}}\left[h_{v} \{\mat{\delta}_{p}
p_{\mathrm{sat}}\vek{\nabla}\varphi\}\right]\right\rangle_{\Box}\mathrm{d}\Omega
- \int_{\Gamma^{\bar{q}}_{\theta}}\left\langle\delta
\theta\,\bar{q}_{\nu}\right\rangle_{\Box}\mathrm{d}\Gamma =0.
\label{eq:hom04}
\end{eqnarray}

In the spirit of the first order homogenization, it is assumed that
the macroscopic temperature and relative humidity vary only linearly
over PUC. This can be achieved by loading its boundary by the
prescribed temperature $\Theta^{\mathrm{hom}}$ and relative humidity
$\Phi^{\mathrm{hom}}$ derived from the uniform macroscopic temperature
$\vek{\nabla}\Theta$ and relative humidity $\vek{\nabla}\Phi$
gradients. In such a case, the local temperature and relative humidity
inside PUC admit the following decomposition
\begin{eqnarray}
\theta(\vek{x}) & = & \Theta(\vek{X}^{0}) + \{\vek{\nabla}
\Theta\}^{\mathrm{\textsf{T}}}\{\vek{x} -
\vek{X}^{0}\}+\theta^{*}(\vek{x}) = \Theta^{\mathrm{hom}}(\vek{x})+\theta^{*}(\vek{x}), \label{eq:hom05} \\
\varphi(\vek{x}) & = & \Phi(\vek{X}^{0}) + \{\vek{\nabla}
\Phi\}^{\mathrm{\textsf{T}}}\{\vek{x} -
\vek{X}^{0}\}+\varphi^{*}(\vek{x})\, =
\Phi^{\mathrm{hom}}(\vek{x}) + \varphi^{*}(\vek{x}),
\label{eq:hom06}
\end{eqnarray}
where $\theta^{*}(\vek{x})$ and $\varphi^{*}(\vek{x})$ are the
fluctuations of local fields superimposed onto linearly varying
quantities $\Theta^{\mathrm{hom}}(\vek{x})$ and
$\Phi^{\mathrm{hom}}(\vek{x})$ . The temperature $\Theta(\vek{X}^{0})$
and the moisture $\Phi(\vek{X}^{0})$ at the reference point
$\vek{X}^{0}$ are introduced to link the local fields to their
macroscopic counterparts. For convenience, the PUC is typically
centered at $\vek{X}^{0}$. Henceforth, the local fluctuations will be
demanded to be periodic, i.e. the same values are enforced on the
opposite sides of a rectangular PUC.  This ensures the scale
transition condition, see e.g.~\cite{Ozdemir:IJNME:2008}, which arises
from upon averaging the micro-temperature gradient over the volume
$\measure{\Omega}$ of PUC
\begin{equation}
\langle\{\vek{\nabla}\theta(\vek{x})\}\rangle_{\Box} = \frac{1}{\measure{\Omega_{\Box}}}\int_{\Omega_{\Box}} \{\vek{\nabla}\theta(\vek{x})\} \de\Omega(\vek{x}) =
\{\vek{\nabla}\Theta\}+
\frac{1}{\measure{\Omega_{\Box}}}\int_{\Omega_{\Box}}\{\vek{\nabla}\theta^{*}(\vek{x})\} \de\Omega(\vek{x}).\label{eq:avggT}
\end{equation}
Hence we demand the contribution of fluctuation fields to disappear
upon volume averaging
\begin{equation}
\langle \{\vek{\nabla}\theta^{*}(\vek{x})\} \rangle_{\Box} = \frac{1}{\measure{\Omega_{\Box}}}\int_{\Omega_{\Box}}
\{\vek{\nabla}\theta^{*}(\vek{x})\} \de\Omega(\vek{x}) =
\frac{1}{\measure{\Omega_{\Box}}}\int_{\Gamma_{\Box}} \theta^*(\vek{x})\{\vek{\nu}(\vek{x})\}
\de\Gamma(\vek{x}) = 0, \label{eq:t*}
\end{equation}
where $\vek{\nu}$ stores the components of the outward unit normal to
the boundary of PUC being directed in opposite directions on
opposite sides of the PUC.

Next, substituting Eq. (\ref{eq:hom05}) into Eq. (\ref{eq:hom04}) and
collecting the terms corresponding to $\delta\Theta^{\mathrm{hom}}$
and $\delta\theta^*$ splits the original problem~\eqref{eq:hom01} into
the homogenized (macro-scale) problem and local sub-scale (meso-scale)
problem. Since details on the derivation of equations driving the
solution on individual scales have already been given in our
proceeding works~\cite{Sykora:2011:AMC,Sykora:JCAM:2011}, we present
only the result pertinent to the macro-scale
\begin{eqnarray}
&&-\underbrace{\int_{\Omega}\left\langle\{\delta
\Theta\}^{\mathrm{\textsf{T}}}
\frac{\mathrm{d}H}{\mathrm{d}\theta}\frac{\mathrm{d}}{\mathrm{d}t}\left(\Theta+\vek{\nabla}\Theta^{\mathrm{\textsf{T}}}\{\vek{x}-\vek{X}^{0}\}\right)
\right\rangle_{\Box}\mathrm{d}\Omega}_{\left(\mat{C}_{\theta\theta}+\mat{C}_{\theta\theta}^{'}\right)\frac{\mathrm{d}\vek{r}_{\theta}}{\mathrm{d}t}}
- \nonumber \\
&& -\underbrace{\int_{\Omega}\left\langle \{\delta\vek{\nabla}\Theta\}^{\mathrm{\textsf{T}}} \{\vek{x}-\vek{X}^{0}\}
\frac{\mathrm{d}H}{\mathrm{d}\theta}\frac{\mathrm{d}}{\mathrm{d}t}\left(\Theta+\vek{\nabla}\Theta^{\mathrm{\textsf{T}}}\{\vek{x}-\vek{X}^{0}\}\right)\right\rangle_{\Box}
\mathrm{d}\Omega}_{\left(\mat{C}_{\theta\theta}^{'}+\mat{C}_{\theta\theta}^{''}\right)\frac{\mathrm{d}\vek{r}_{\theta}}{\mathrm{d}t}}
- \nonumber \\
&& -\underbrace{\int_{\Omega}\left\langle\{\delta\vek{\nabla}
\Theta \}^{\mathrm{\textsf{T}}}\left[\{\mat{\lambda}\vek{\nabla}
\Theta\}+h_{v}
\left\{\mat{\delta}_{p}\varphi\frac{\mathrm{d}p_{\mathrm{sat}}}{\mathrm{d}\theta}\vek\nabla\Theta\right\}\right]\right\rangle_{\Box}\mathrm{d}\Omega}_{\mat{K}_{\theta\theta}r_{\theta}} -\nonumber \\
&& -\underbrace{\int_{\Omega}\left\langle\{\delta \vek{\nabla}
\Theta \}^{\mathrm{\textsf{T}}}\left[h_{v} \{\mat{\delta}_{p}
p_{\mathrm{sat}}\vek{\nabla}\Phi\}\right]\right\rangle_{\Box}\mathrm{d}\Omega}_{\mat{K}_{\theta\varphi}r_{\varphi}}
+ \underbrace{\int_{\Gamma^{\bar{q}}_{\theta}}\left\langle\{\delta
\Theta\}^{\mathrm{\textsf{T}}}
\,\bar{q}_{\nu}\right\rangle_{\Box}\mathrm{d}\Gamma}_{\vek{q}_{\mathrm{ext}}}
=0.\label{eq:hom10}
\end{eqnarray}
to identify the solution dependence on the actual size of PUC through
the second term in the integral~\eqref{eq:hom10}.

An analogous approach can be applied also to the moisture
transport equation~\eqref{eq:hom01m} to arrive, after classical
finite element discretization, into a discretized system of
coupled macroscopic heat and moisture equations
\begin{eqnarray}
\mat{K}_{\theta\theta}r_{\theta}+\mat{K}_{\theta\varphi}r_{\varphi}+(\mat{C}_{\theta\theta}+2\mat{C}_{\theta\theta}^{'}+\mat{C}_{\theta\theta}^{''})\frac{\mathrm{d}\vek{r}_{\theta}}{\mathrm{d}t}
& = & \vek{q}_{\mathrm{ext}}, \label{eq:hom11}\\
\mat{K}_{\varphi\theta}r_{\theta}+\mat{K}_{\varphi\varphi}r_{\varphi}+(\mat{C}_{\varphi\varphi}+2\mat{C}_{\varphi\varphi}^{'}+\mat{C}_{\varphi\varphi}^{''})\frac{\mathrm{d}\vek{r}_{\varphi}}{\mathrm{d}t}
& = & \vek{g}_{\mathrm{ext}}, \label{eq:hom12}
\end{eqnarray}
which have to be properly integrated in the time domain adopting for
example the Crank-Nicolson integration scheme. Details on the
numerical implementation are available in~\cite{Sykora:2010}.  The
homogenized matrices in Eqs.~\eqref{eq:hom11} and~\eqref{eq:hom12}
follow directly from the meso-scale solution for a given macroscopic
time increment. Because of a strong non-linearity arising on both
scales the two problems (macro-meso) must be solved iteratively by the
Newton-Raphson method,
see~\cite{Ozdemir:IJNME:2008,Larsson:2010:IJMNE,Sykora:JCAM:2011} for
further reference.

\subsection{Numerical examples}
%%%%%%%%%%%%%%%%%%%%%%%%%%%%%%%%%%%%%%%%%%%%%%%%%%%%%%%%%%%%%%%%%%%%%%%%%%%%%%%%%
A fully coupled multi-scale analysis of a two-dimensional segment of
Charles Bridge in Prague has been carried out
in~\cite{Sykora:JCAM:2011} with emphases on parallel computing.  This
study will be extended herein by quantifying the influence of the
boundary, loading and initial conditions when moving down from the
macro-scale to the meso-scale.

\subsubsection{Boundary conditions}\label{sec:BC}
%%%%%%%%%%%%%%%%%%%%%%%%%%%%%%%%%%%%%%%%%%%%%%%%%%%%%%%%%%%%%%%%%%%%%%%%%%%%%%%%%
To begin, recall Eq.~\eqref{eq:t*}
\begin{equation}
\langle \{\vek{\nabla}\theta^{*}(\vek{x})\} \rangle_{\Box} =
\frac{1}{\measure{\Omega_{\Box}}}\int_{\Gamma_{\Box}} \theta^*(\vek{x})\{\vek{\nu}(\vek{x})\}
\de\Gamma(\vek{x}) = 0, \label{eq:t2*}
\end{equation}
being satisfied providing either the fluctuation part of the
temperature field equals zero or the periodic boundary conditions,
i.e. the same values of $\theta^*$ on opposite sides of a rectangular
PUC, are enforced on $\Gamma_{\Box}$.

%%%%%%%%%%%%%%%%%%%%%%%%%%%%%%%%%%%%%%%%%%%%%%%%%%%%%%%%%%%%%%%%%%%%%%%%%%%%%%%%%%%%%%%%%%%%%
\begin{figure} [ht!]
\begin{center}
\begin{tabular}{c}
\includegraphics[width=70mm,keepaspectratio]{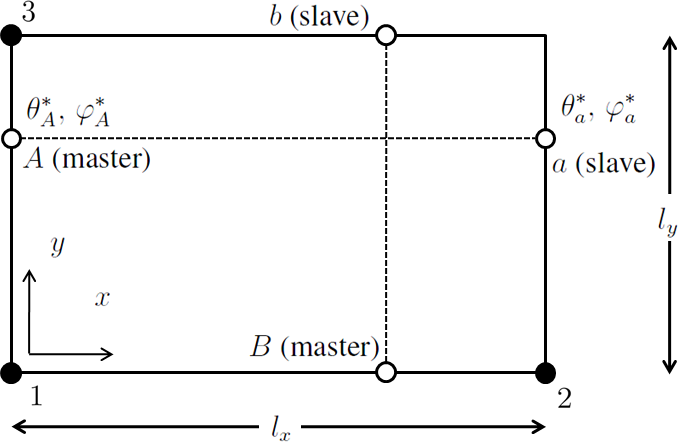}
\end{tabular}
\caption{Periodic boundary conditions} \label{fig:pbc}
\end{center}
\end{figure}
%%%%%%%%%%%%%%%%%%%%%%%%%%%%%%%%%%%%%%%%%%%%%%%%%%%%%%%%%%%%%%%%%%%%%%%%%%%%%%%%%%%%%%%%%%%%%

Such conditions are easy to impose if searching the solution in terms
of the fluctuation part of the temperature or moisture fields. The
macroscopic constant gradients $\{\vek{\nabla}\Theta\}$ and
$\{\vek{\nabla} \Phi\}$, see Figure~\ref{fig:puc}(a), are then
directly used to load the unit cell~\cite{Sykora:2011:AMC}. However,
this is mostly not possible with the application of commercial
codes. In such a case, the constant gradients are introduced by
enforcing a linear variation of the homogeneous part of local fields
$\Theta^{\mathrm{hom}}(\vek{x}), \Phi^{\mathrm{hom}}(\vek{x})$. This
is achieved by prescribing directly the Dirichlet boundary conditions
along all edges of PUC,
i.e. $\theta(\vek{x})=\Theta^{\mathrm{hom}}(\vek{x}),
\varphi(\vek{x})=\Phi^{\mathrm{hom}}(\vek{x})$ (see
Figure~\ref{fig:puc}(b)) are specified on the boundary $\Gamma_{\Box}$
of PUC. The periodic boundary conditions are then prescribed with the
help of multi-point constraints.

In doing so, observe in Figure~\ref{fig:pbc} that for a pair of points
(e.g. $A$ - master and $a$ - slave) located on the opposite sides of
PUC the following relations hold:
\begin{equation}
\theta_A = \left( \frac{\partial\Theta}{\partial y} \right) y_A + \theta^*_A +
\Theta(\vek{X}_0), \quad \theta_a = \left( \frac{\partial\theta}{\partial x}
\right) L + \left( \frac{\partial\Theta}{\partial y} \right) y_a +
\theta^*_a + \Theta(\vek{X}_0).\label{eq:local_t-sifel}
\end{equation}
Taking into account the fact that the fluctuation field $\theta^*$
satisfies the periodicity condition
\begin{equation}
\theta^*_a = \theta^*_A,\label{eq:periodic-sifel}
\end{equation}
and subtracting corresponding terms on the opposite edges, we finally
obtain (compare with~\cite{Ozdemir:IJNME:2008})
\begin{eqnarray}
\left( \frac{\partial\Theta}{\partial x} \right) L  & = & \theta_a - \theta_A = \theta_2 - \theta_1,\nonumber \\
\left( \frac{\partial\Theta}{\partial y} \right) H  & = & \theta_b - \theta_B =
\theta_3 - \theta_1,\label{eq:constr-sifel}
\end{eqnarray}
where $\theta_1, \theta_2$ and $\theta_3$ are the temperatures at the
control points 1,2 and 3 seen in Figure~\ref{fig:pbc}. The same
conditions apply to the relative humidity $\varphi$ as well.

\subsubsection{Loading and initial conditions}
%%%%%%%%%%%%%%%%%%%%%%%%%%%%%%%%%%%%%%%%%%%%%%%%%%%%%%%%%%%%%%%%%%%%%%%%%%%%%%%%%
An illustrative example presented here considers a finite element
mesh consisting of $108$ macro-elements each representing a single
meso-problem with assigned periodic boundary conditions.  Its geometry
together with the specific macroscopic loading conditions are shown in
Figure~\ref{fig:scheme}.

The initial temperature $\theta_{\mathrm{in}} = 14$ [$^{\circ}$C] and
the moisture $\varphi_{\mathrm{in}} = 0.5$ [-] were assigned to the
whole domain. The following boundary and loading conditions were
imposed: the left boundary of the domain was submitted to exterior
loading conditions $\theta_{\mathrm{ext}} = 5$ [$^{\circ}$C] and
$\varphi_{\mathrm{ext}} = 0.5$ [-], while the opposite side was
submitted to interior loading conditions $\theta_{\mathrm{int}} = 24$
[$^{\circ}$C] and $\varphi_{\mathrm{int}} = 0.8$ [-]. Zero flux
boundary conditions were assumed for horizontal edges.

%%%%%%%%%%%%%%%%%%%%%%%%%%%%%%%%%%%%%%%%%%%%%%%%%%%%%%%%%%%%%%%%%
\begin{table}[ht!]
\begin{center}
\begin{tabular}{lllcc}
\multicolumn{3}{l}{parameter} & brick & mortar \\
\hline
$w_{f}$ &  $\mathrm{[kgm^{-3}]}$ &free water saturation & 229.30 & 160.00 \\
$w_{\mathrm{80}}$  & $\mathrm{[kgm^{-3}]}$ & water content at $\varphi=0.8$ [-] & 141.68 & 22.72 \\
$\lambda_{\mathrm{0}}$ & $\mathrm{[Wm^{-1}K^{-1}]}$ & thermal conductivity & 0.25 & 0.45 \\
$b_{\mathrm{tcs}}$  & $\mathrm{[-]}$ & thermal conductivity supplement & 10 & 9  \\
$\rho_{s}$ & $\mathrm{[kgm^{-3}]}$ &  bulk density & 1690 & 1670 \\
$\mu$   & $\mathrm{[-]}$ &water vapor diffusion resistance & 16.80 & 9.63 \\
$A$ & $\mathrm{[kgm^{-2}s^{-0.5}]}$ &  water absorption coefficient & 0.51 & 0.82 \\
$c_s$ & $\mathrm{[Jkg^{-1}K^{-1}]}$ &specific heat capacity & 840 & 1000 \\
\hline
\end{tabular}
\caption{Material parameters of individual phases}
\label{tab:matpar}
\end{center}
\end{table}
%%%%%%%%%%%%%%%%%%%%%%%%%%%%%%%%%%%%%%%%%%%%%%%%%%%%%%%%%%%%%%%%%

The material parameters in Table~\ref{tab:matpar} were obtained from a
set of experimental measurements providing the hygric and thermal
properties of mortars and bricks/stones, which have been used in the
reconstruction works of historical buildings in the Czech Republic,
see~\cite{Sykora:2011:AMC,Pavlikova:NSBP:2008}.

With reference to Section~\ref{sec:BC} we consider two types of
loading conditions. Figure~\ref{fig:puc}(a) assumes the loading in terms
of macroscopic temperature and moisture gradients to provide local
fluctuations upon submitting decompositions~\eqref{eq:hom05}
and~\eqref{eq:hom06} into Hill's averaging condition (consider for
simplicity a steady-state heat conduction problem only)
\begin{equation}
\langle\{\delta\vek{\nabla}\theta\}\trn\{q\}\rangle = 0,
\end{equation}
to get
\begin{equation}
\langle\{\delta\vek{\nabla}\theta^*\}\trn\emtrx{\lambda}\{\vek{\nabla}\theta^*\}\rangle
=
-\langle\{\delta\vek{\nabla}\theta^*\}\trn\emtrx{\lambda}\{\vek{\nabla}\Theta\}\rangle.
\end{equation}
The periodic boundary conditions are then easily introduced by
prescribing the same code numbers, in the FEM language, to the
homologous nodes on the opposite faces of the rectangular unit
cell. Figure~\ref{fig:puc}(b) then presents fully prescribed Dirichlet
boundary conditions applicable with commercial codes when searching
the solution directly in terms of the local fields
$\theta,\,\varphi$. The periodic boundary conditions are then enforced
indirectly using Eq.~\eqref{eq:constr-sifel}.

%%%%%%%%%%%%%%%%%%%%%%%%%%%%%%%%%%%%%%%%%%%%%%%%%%%%%%%%%%%%%%%%%
\begin{figure} [ht!]
\begin{center}
\begin{tabular}{c@{\hspace{2mm}}c@{\hspace{2mm}}c}
\includegraphics*[width=47mm,keepaspectratio]{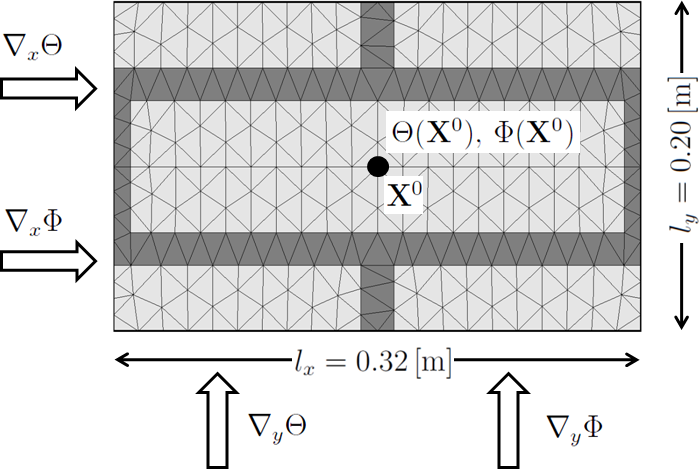}&
\includegraphics*[width=47mm,keepaspectratio]{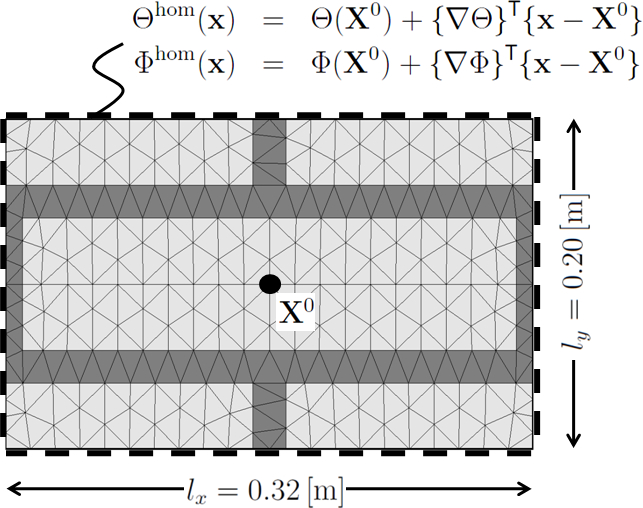}&
\includegraphics*[width=47mm,keepaspectratio]{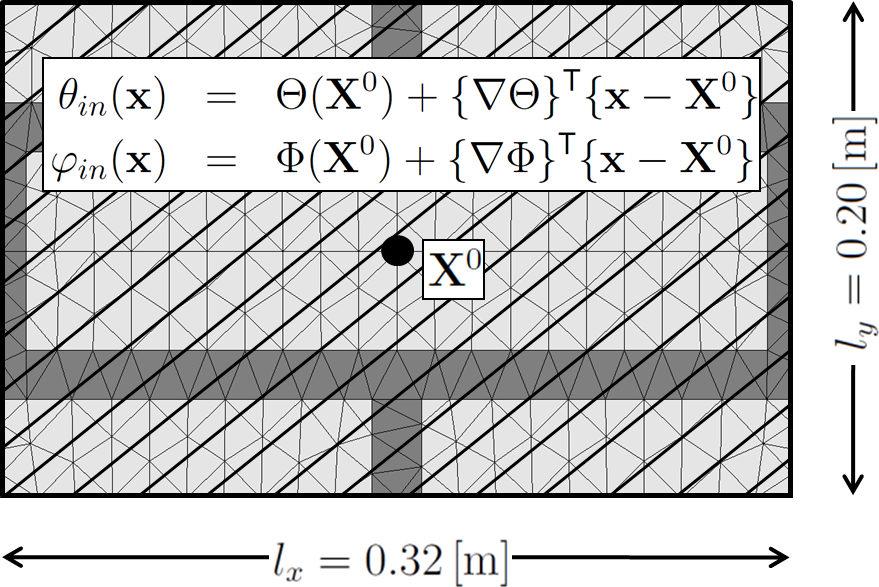}\\
(a)&(b)&(c)
\end{tabular}
\end{center}
\caption{Different loading schemes of PUC on meso-scale:
(a) prescribed macroscopic gradients (loading conditions I, labeled as NST$^1$),
(b) prescribed macroscopic temperatures and relative humidities (loading conditions II, labeled as NST$^2$);
(c) Assumed initial conditions} \label{fig:puc}
\end{figure}
%%%%%%%%%%%%%%%%%%%%%%%%%%%%%%%%%%%%%%%%%%%%%%%%%%%%%%%%%%%%%%%%%

Apart from loading conditions, the two types of initial conditions
seen in Figure~\ref{fig:puc}(c) are also examined. In particular,
either a linear variation of the homogeneous part of local fields
$\theta_{\mathrm{in}}(\vek{x})=\Theta^{\hom}(\vek{x}),\,\varphi_{\mathrm{in}}(\vek{x})=\Phi^{\hom}(\vek{x})$
is assumed or the macroscopic temperature
$\theta_{\mathrm{in}}(\vek{x})=\Theta(\vek{X}^0)$ and moisture
$\varphi_{\mathrm{in}}(\vek{x})=\Phi(\vek{X}^0)$ calculated at the
reference point $\vek{X}^0$ at the end of the current macroscopic time
step are assigned to the whole PUC. Note that the former one arise
naturally from Eqs.~\eqref{eq:hom05} and~\eqref{eq:hom06} when setting
the fluctuation fields equal to zero at time $t=0$.

\subsubsection*{Meso-scale analysis}
%%%%%%%%%%%%%%%%%%%%%%%%%%%%%%%%%%%%%%%%%%%%%%%%%%%%%%%%%%%%%%%%% 
In the first example attention is dedicated to the solution of a
coupled transient heat and moisture problem solely on the meso-scale.
First, the regular periodic unit cell, PUC in Figure~\ref{fig:scheme},
was loaded by the highest temperature and moisture gradients obtained
in the course of actual multi-scale analysis discussed later in this
section. The following numerical values were considered; initial
conditions ($\Theta(\vek{X}^{0})=17\,\mathrm{[^{\circ}C]}$,
$\Phi(\vek{X}^{0})=0.15, 0.55, 0.85\,\mathrm{[-]}$) and boundary
conditions ($\{\vek{\nabla}
\Theta\}=\{\nabla_{x}\Theta,\nabla_{y}\Theta\}=\{5.0,1.0\}\,\mathrm{[^{\circ}Cm^{-1}]}$,
$\{\vek{\nabla}
\Phi\}=\{\nabla_{x}\Phi,\nabla_{y}\Phi\}=\{0.1,0.05\}\,\mathrm{[m^{-1}]}$).

Figure~\ref{fig:meso-evol} shows evolution of local temperature and
moisture fields along the horizontal centerline of PUC as a
function of time derived from two types of loading conditions.
Periodic boundary conditions and linearly varying initial
conditions with $\Phi(\vek{X}^{0})=0.55\,\mathrm{[-]}$ were
considered. It is shown that for the adopted extreme gradients the
steady state solution, plotted as solid lines, is reached in about
15 hours irrespective of loading conditions.

%%%%%%%%%%%%%%%%%%%%%%%%%%%%%%%%%%%%%%%%%%%%%%%%%%%%%%%%%%%%%%%%%
\begin{figure}[ht!]
\begin{center}
\begin{tabular}{c@{\hspace{5mm}}c}
\includegraphics*[width=70mm,keepaspectratio]{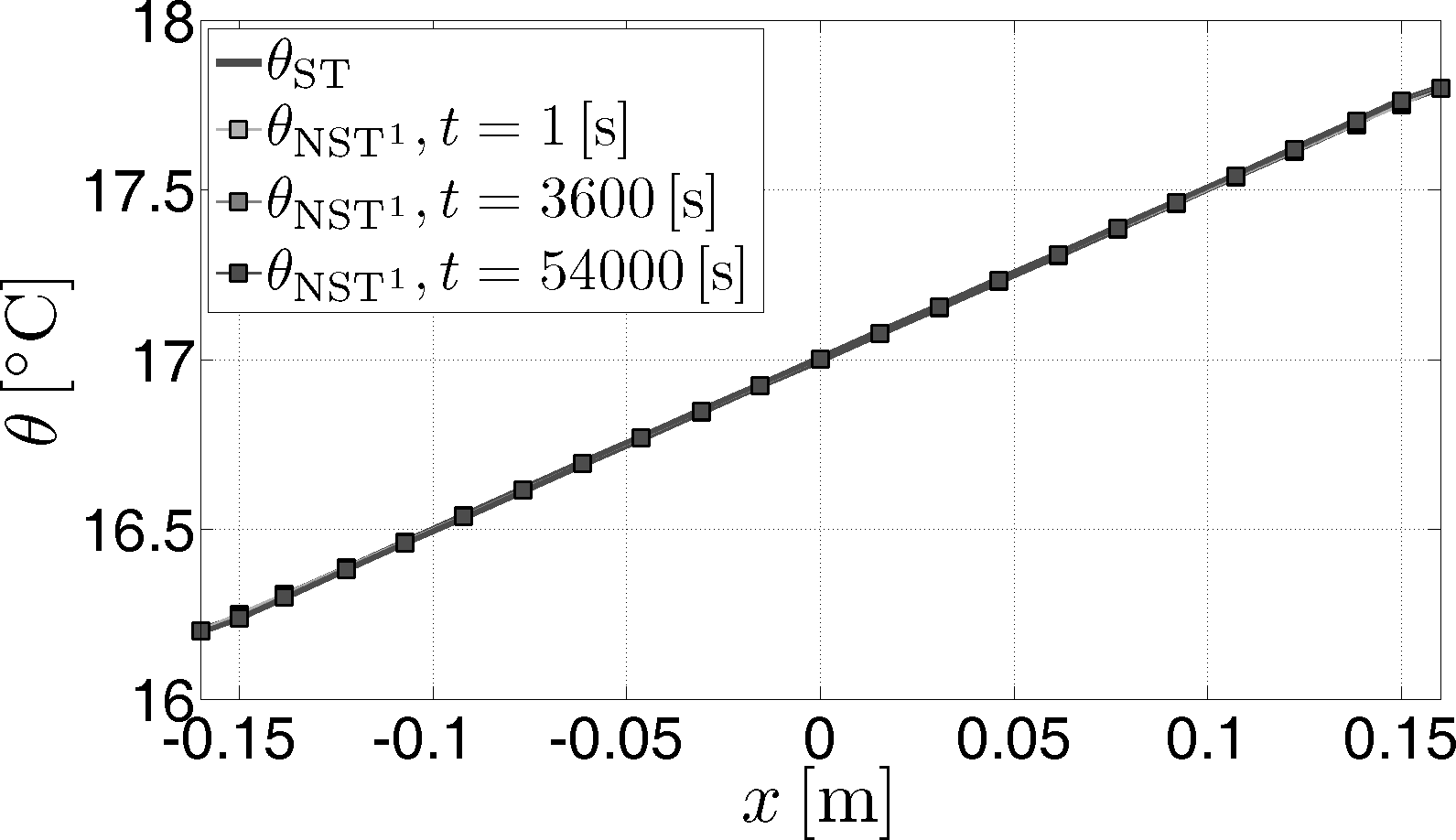}&
\includegraphics*[width=70mm,keepaspectratio]{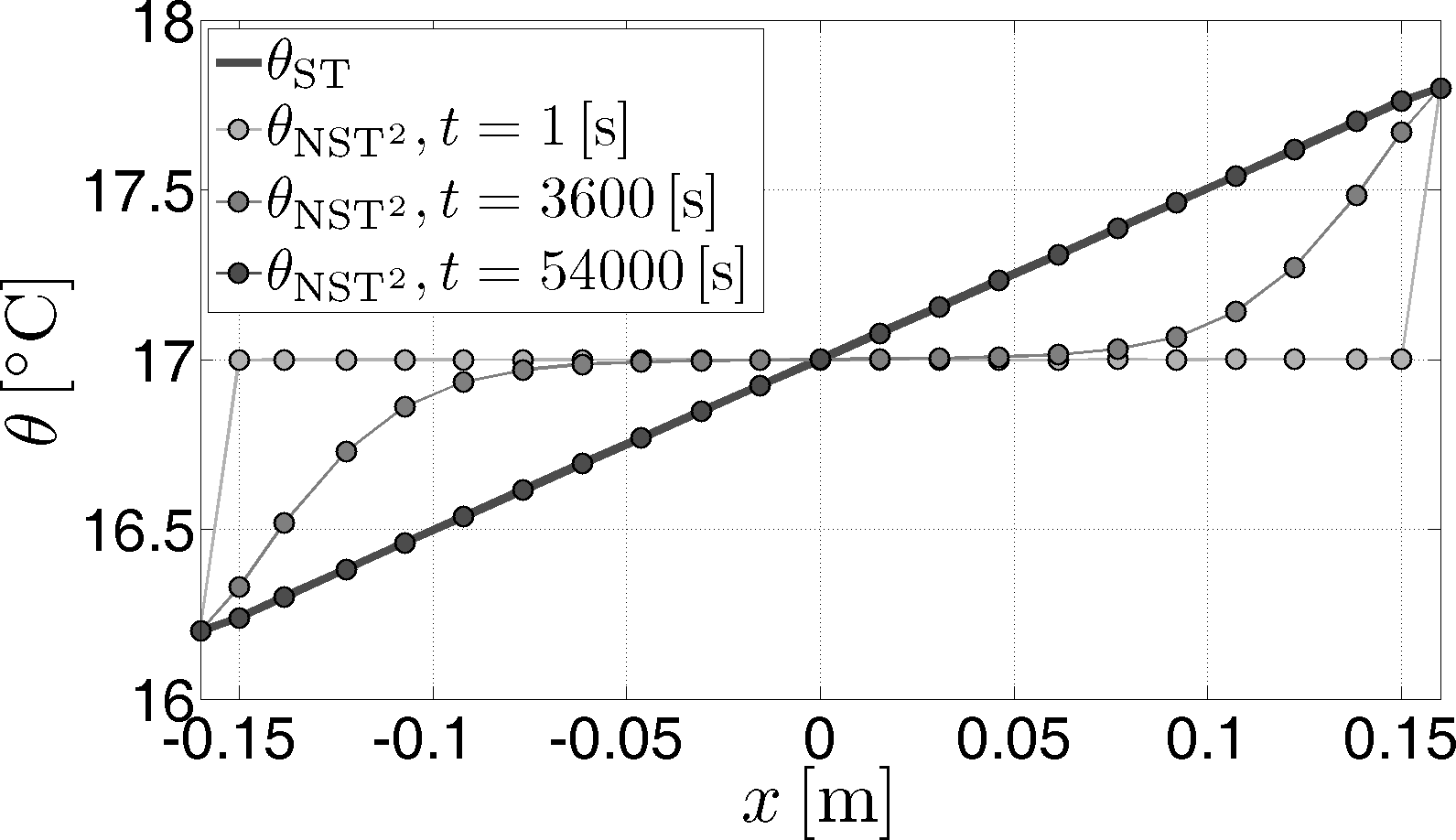}\\
(a)&(b)\\
\includegraphics*[width=70mm,keepaspectratio]{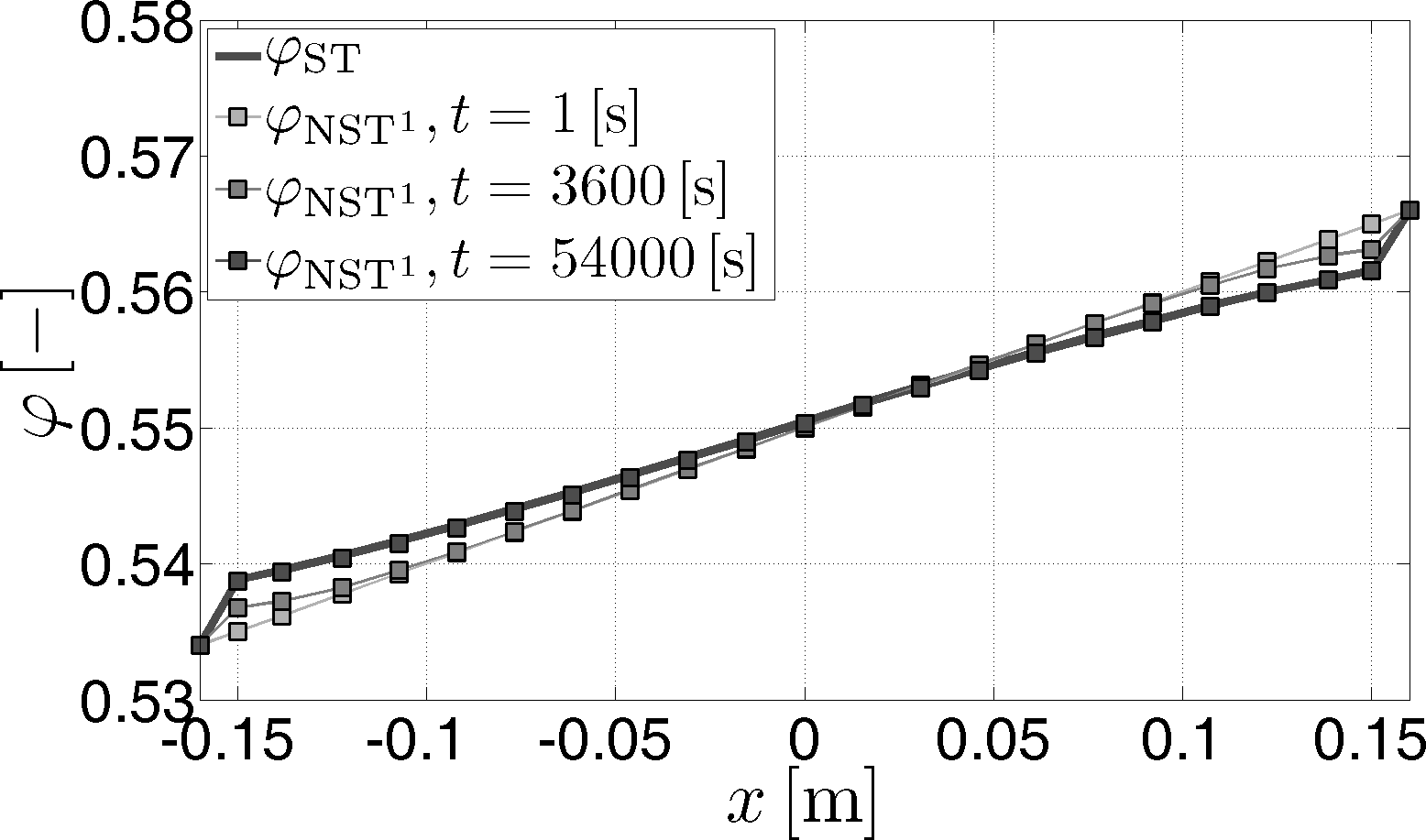}&
\includegraphics*[width=70mm,keepaspectratio]{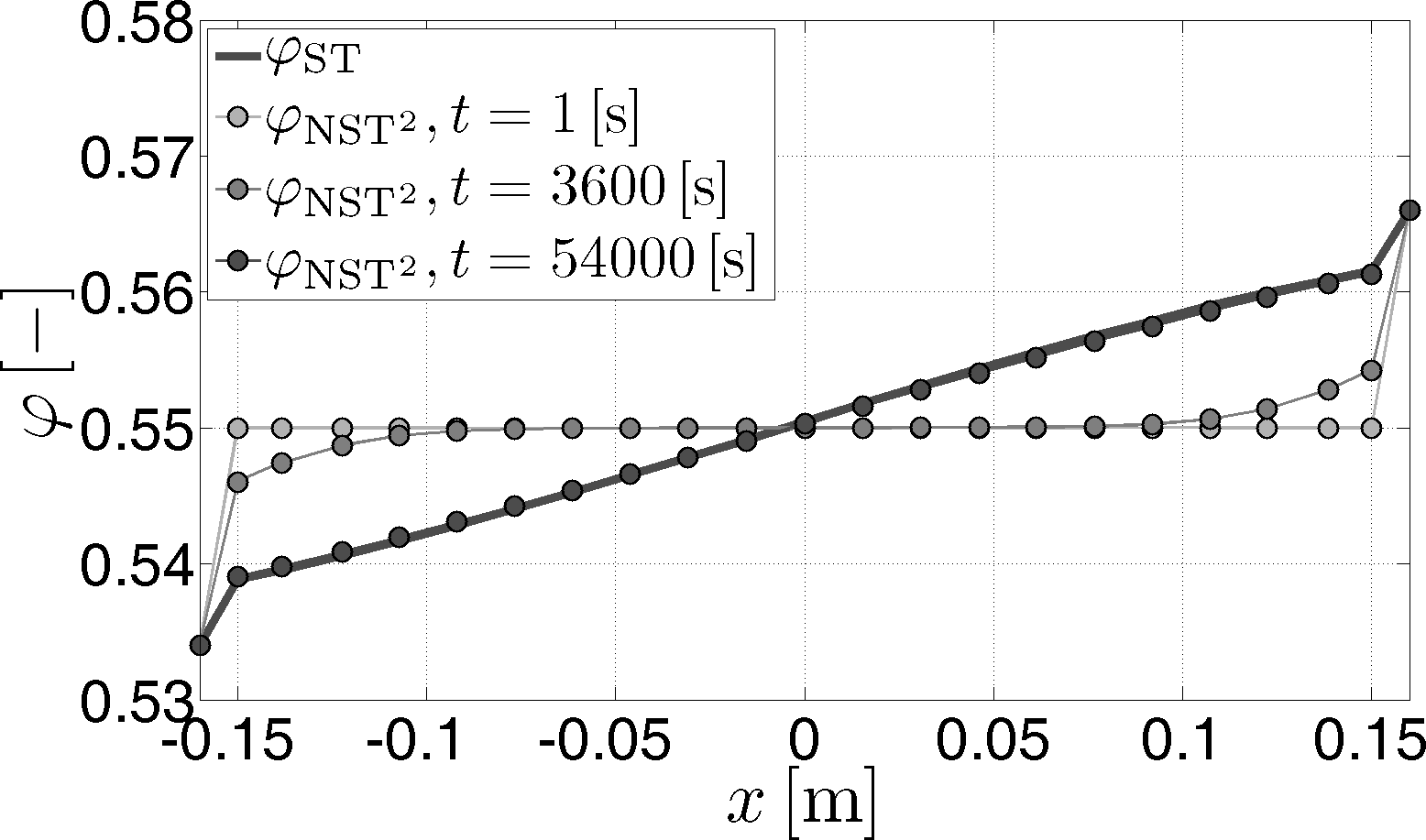}\\
(c)&(d)
\end{tabular}
\end{center}
\caption{Evolution of local fields (PUC): (a,b) temperature, (c,d) moisture;
(a,c) Loading conditions I (Figure~\ref{fig:puc}(a)),
(b,d) Loading conditions II (Figure~\ref{fig:puc}(b))}
\label{fig:meso-evol}
%\end{figure}
%%%%%%%%%%%%%%%%%%%%%%%%%%%%%%%%%%%%%%%%%%%%%%%%%%%%%%%%%%%%%%%%%
%%%%%%%%%%%%%%%%%%%%%%%%%%%%%%%%%%%%%%%%%%%%%%%%%%%%%%%%%%%%%%%%%
%\begin{figure} [ht!]
\begin{center}
\begin{tabular}{c@{\hspace{5mm}}c}
\includegraphics*[width=70mm,keepaspectratio]{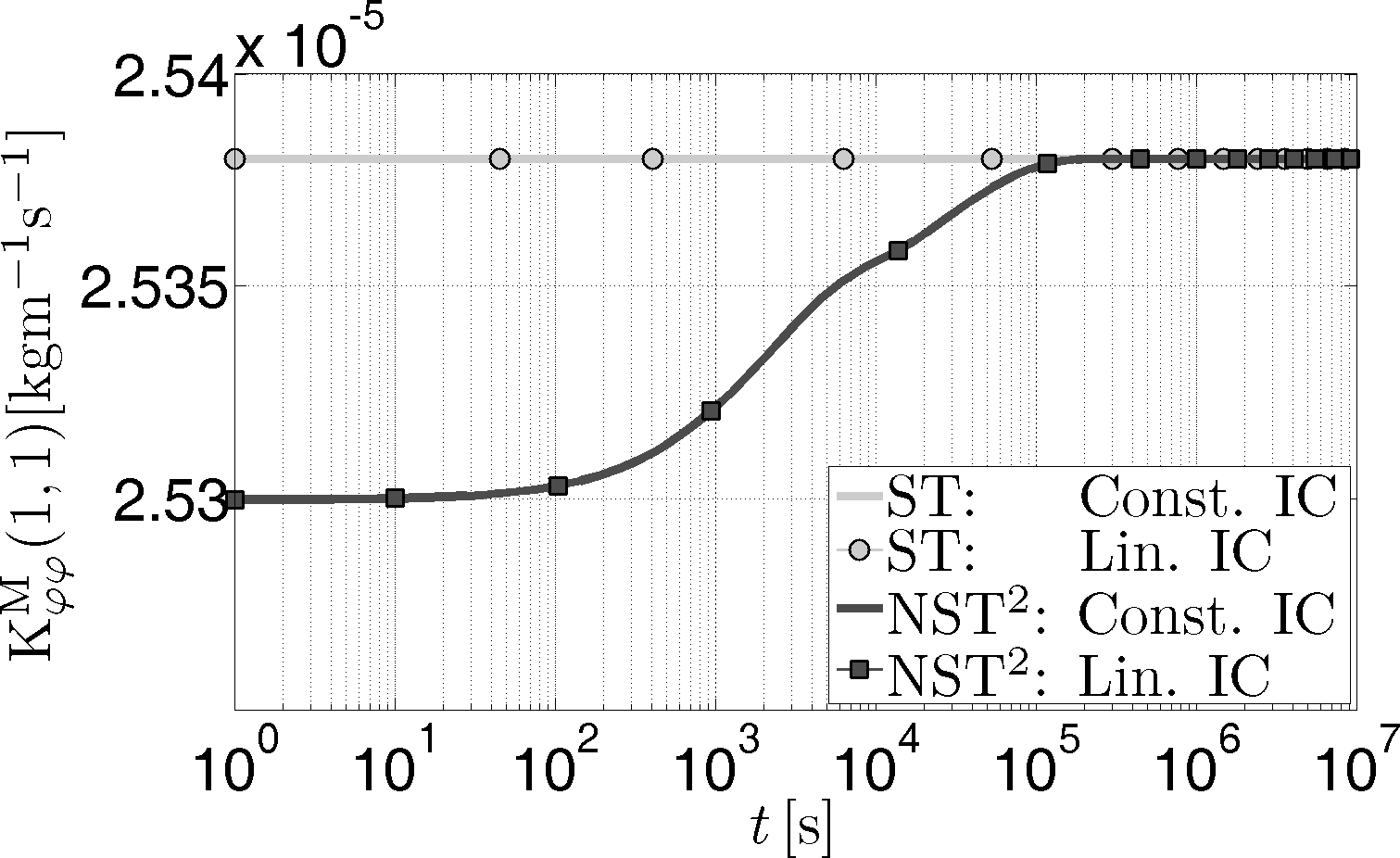}&
\includegraphics*[width=70mm,keepaspectratio]{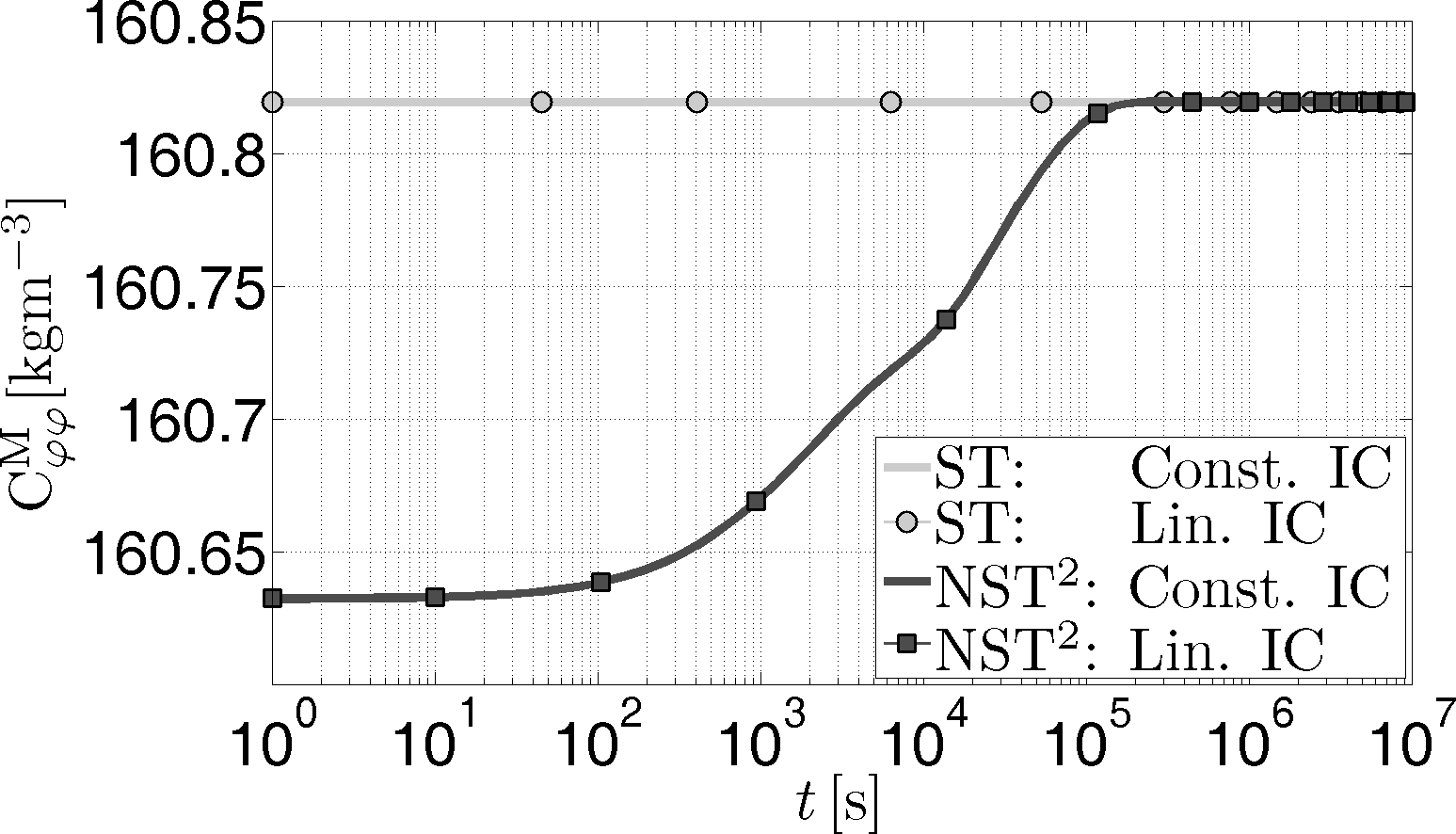}\\
(a)&(b)
\end{tabular}
\end{center}
\caption{Influence of initial conditions (constant or linearly varying
  homogeneous part of local fields); Evolution of effective moisture
  terms as a function of time: (a) effective conductivity term, (b)
  effective storage term}
\label{fig:meso-IC}
\end{figure}
%%%%%%%%%%%%%%%%%%%%%%%%%%%%%%%%%%%%%%%%%%%%%%%%%%%%%%%%%%%%%%%%%

Further to this subject, we also suggest invariance of the solution to
the assumed initial conditions as depicted in Figure~\ref{fig:meso-IC}.
This becomes evident once realizing a nonlinearity of
Eqs.~\eqref{eq:hom01} and~\eqref{eq:hom02} taken into account through
the application of Newton-Raphson iteration scheme. Clearly, the
initial difference in the solution error attributed to the difference
in initial conditions is wiped out already in the first load (time)
increment upon arriving at equilibrium.

The influence of the degree of material nonlinearity of the present
constitutive model is partially seen in Figure~\ref{fig:KC-evol} showing
evolution of the selected effective material properties as a function
of time for three values of initial macroscopic moisture
$\Phi(\vek{X}^{0})$.  Functional dependence of some material
parameters on moisture is plotted in Figure~\ref{fig:model} for
illustration. The complete set is available,
e.g. in~\cite{Sykora:2010,Sejnoha:2012}. It is evident that for
effective properties the time to reach the steady state solution is
considerably shorter than for local temperatures and humidities,
recall Figure~\ref{fig:meso-evol}. It is even more important to realize
that the difference between effective steady state parameters and
effective parameters withing a transient regime is essentially
negligible. This can be attributed to a relatively small difference
between local fields pertinent to steady state and various stages of
transient solutions as evident in Figure~\ref{fig:meso-evol}. In other
words, the material nonlinearity observed in Figure~\ref{fig:model} does
not play in this case a significant role, thus promoting the steady
state solution from the macroscopic point of view as sufficiently
accurate.

%%%%%%%%%%%%%%%%%%%%%%%%%%%%%%%%%%%%%%%%%%%%%%%%%%%%%%%%%%%%%%%%%
\begin{figure} [ht!]
\begin{center}
\begin{tabular}{c@{\hspace{3mm}}c@{\hspace{3mm}}c}
\includegraphics*[width=50mm,keepaspectratio]{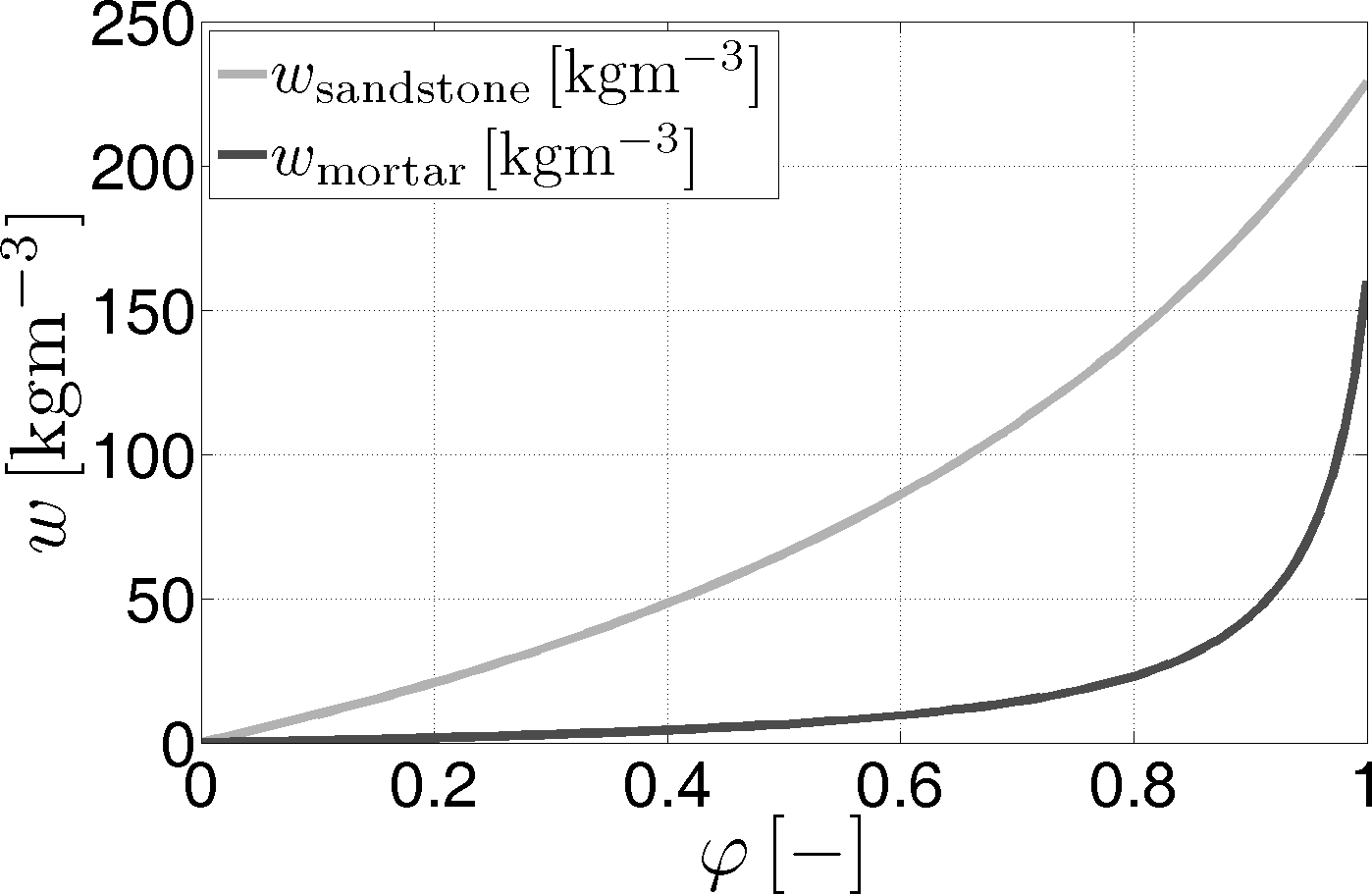}&
\includegraphics*[width=50mm,keepaspectratio]{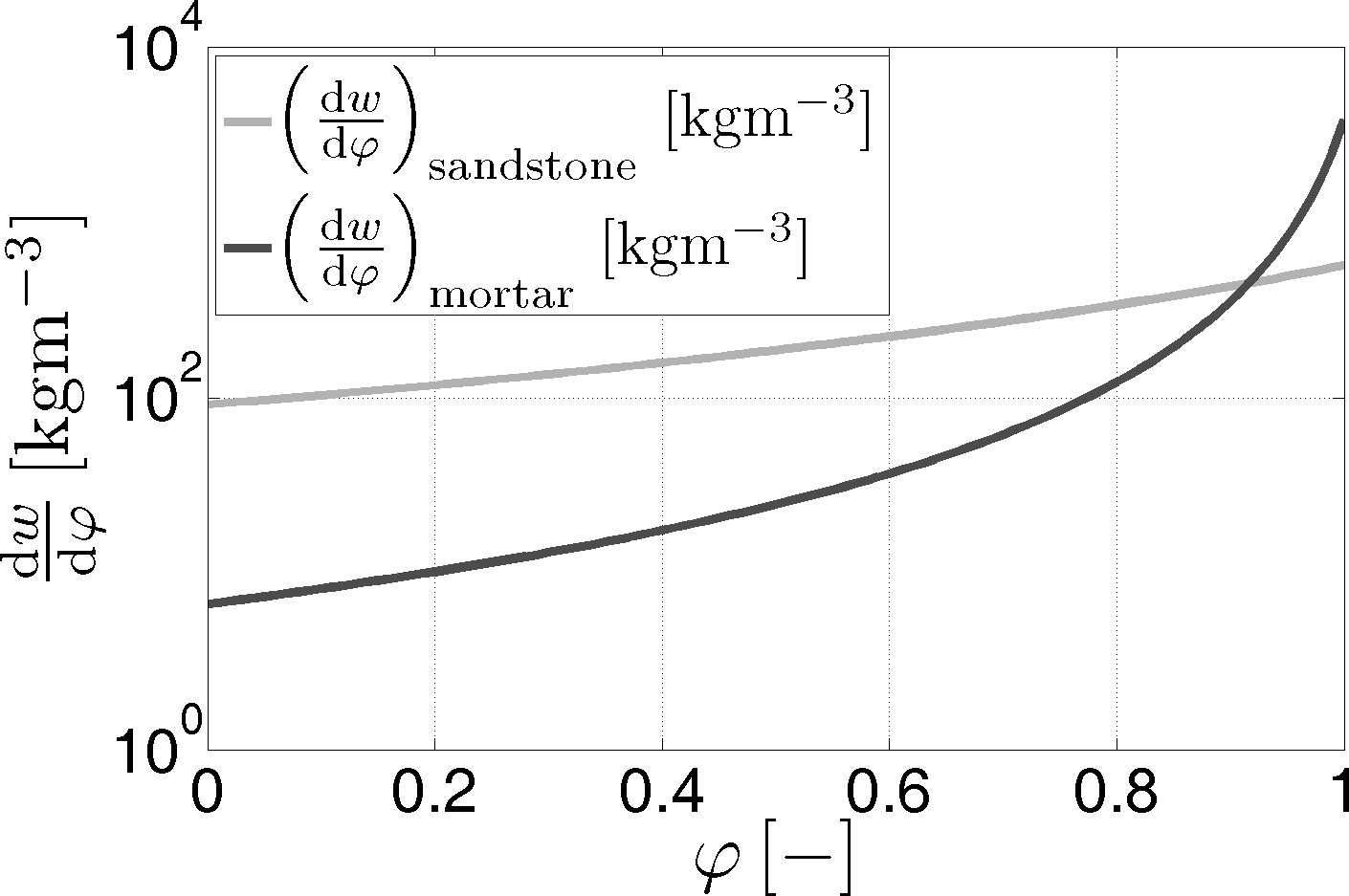}&
\includegraphics*[width=50mm,keepaspectratio]{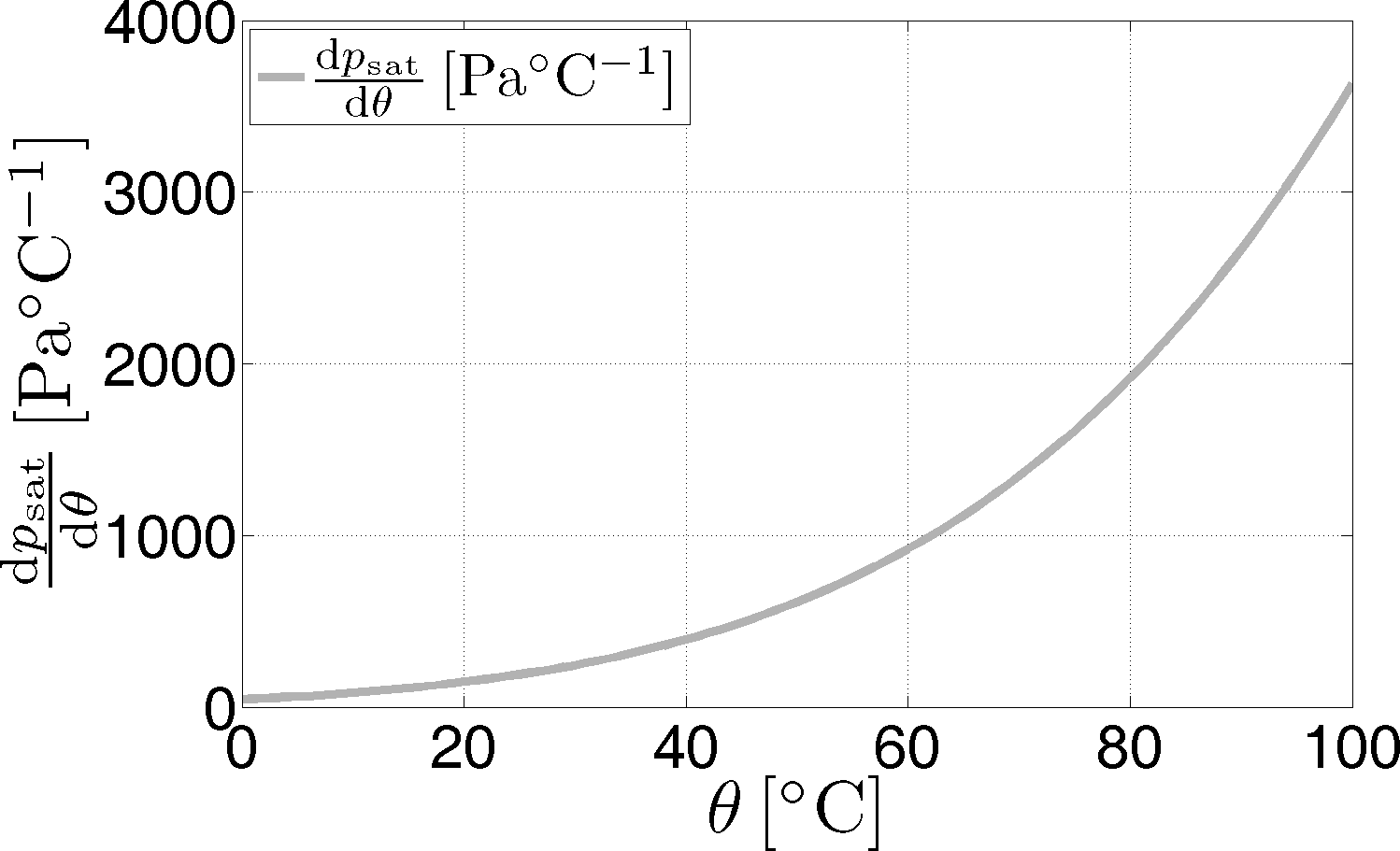}\\
(a)&(b)&(c)
\end{tabular}
\end{center}
\caption{
(a) Variation of water content as a function of relative humidity,
(b) Variation of phase moisture storage function as a function of relative humidity,
(c) Variation of slope of saturation pressure as a function of temperature
}\label{fig:model}
%\end{figure}
%%%%%%%%%%%%%%%%%%%%%%%%%%%%%%%%%%%%%%%%%%%%%%%%%%%%%%%%%%%%%%%%%
%%%%%%%%%%%%%%%%%%%%%%%%%%%%%%%%%%%%%%%%%%%%%%%%%%%%%%%%%%%%%%%%%
%\begin{figure} [ht!]
\begin{center}
\begin{tabular}{c@{\hspace{5mm}}c}
\includegraphics*[width=70mm,keepaspectratio]{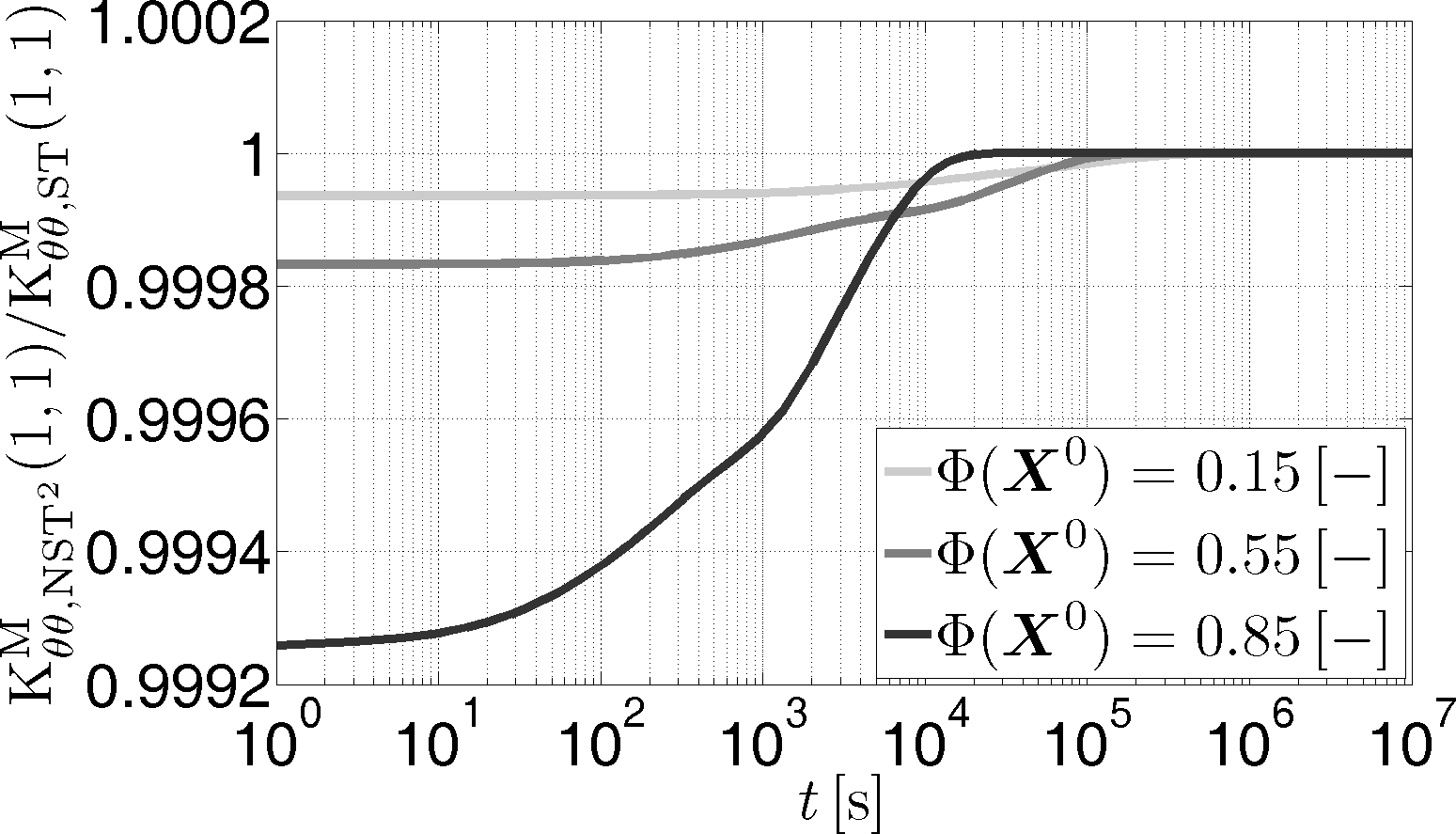}&
\includegraphics*[width=70mm,keepaspectratio]{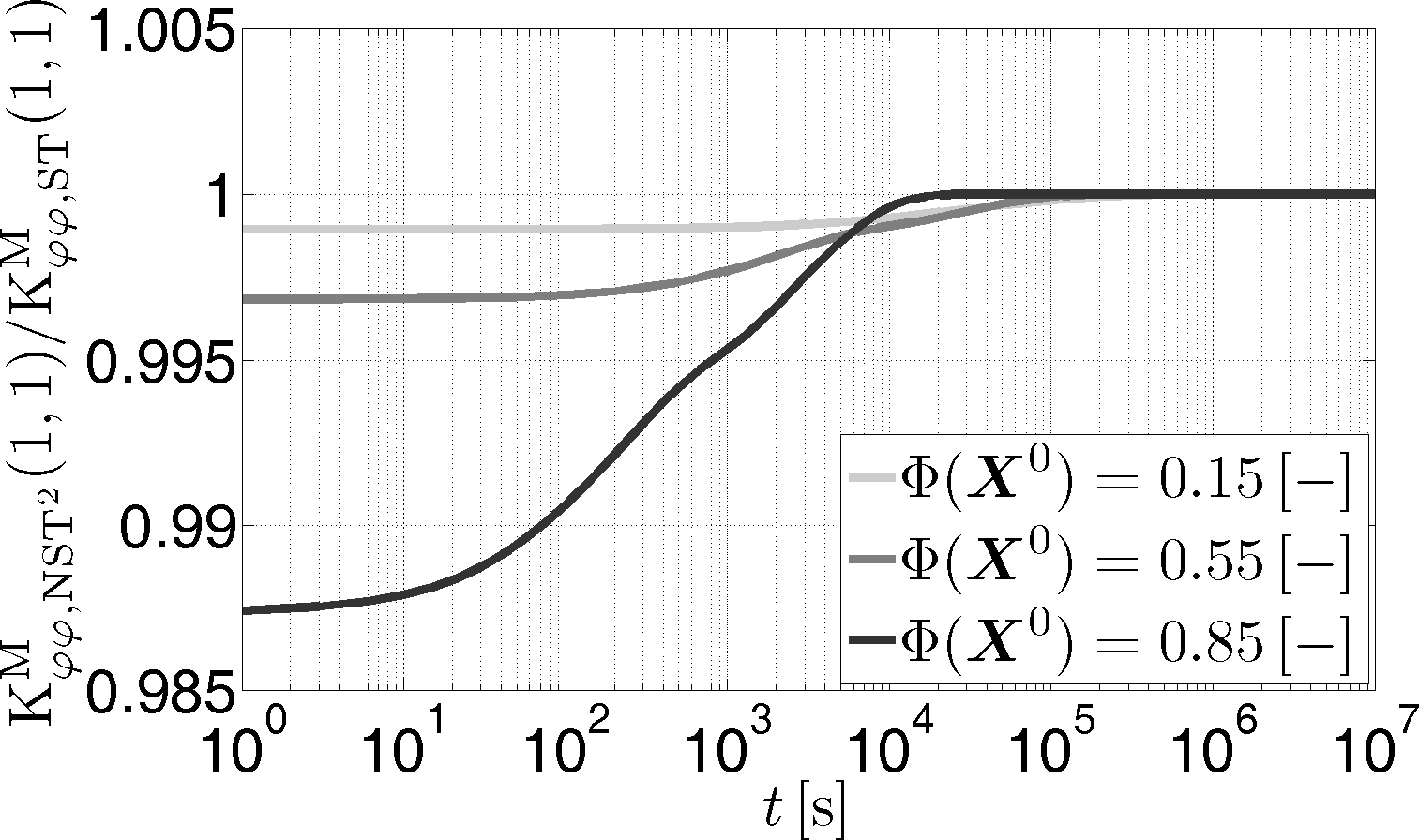}\\
(a)&(b)\\
\includegraphics*[width=70mm,keepaspectratio]{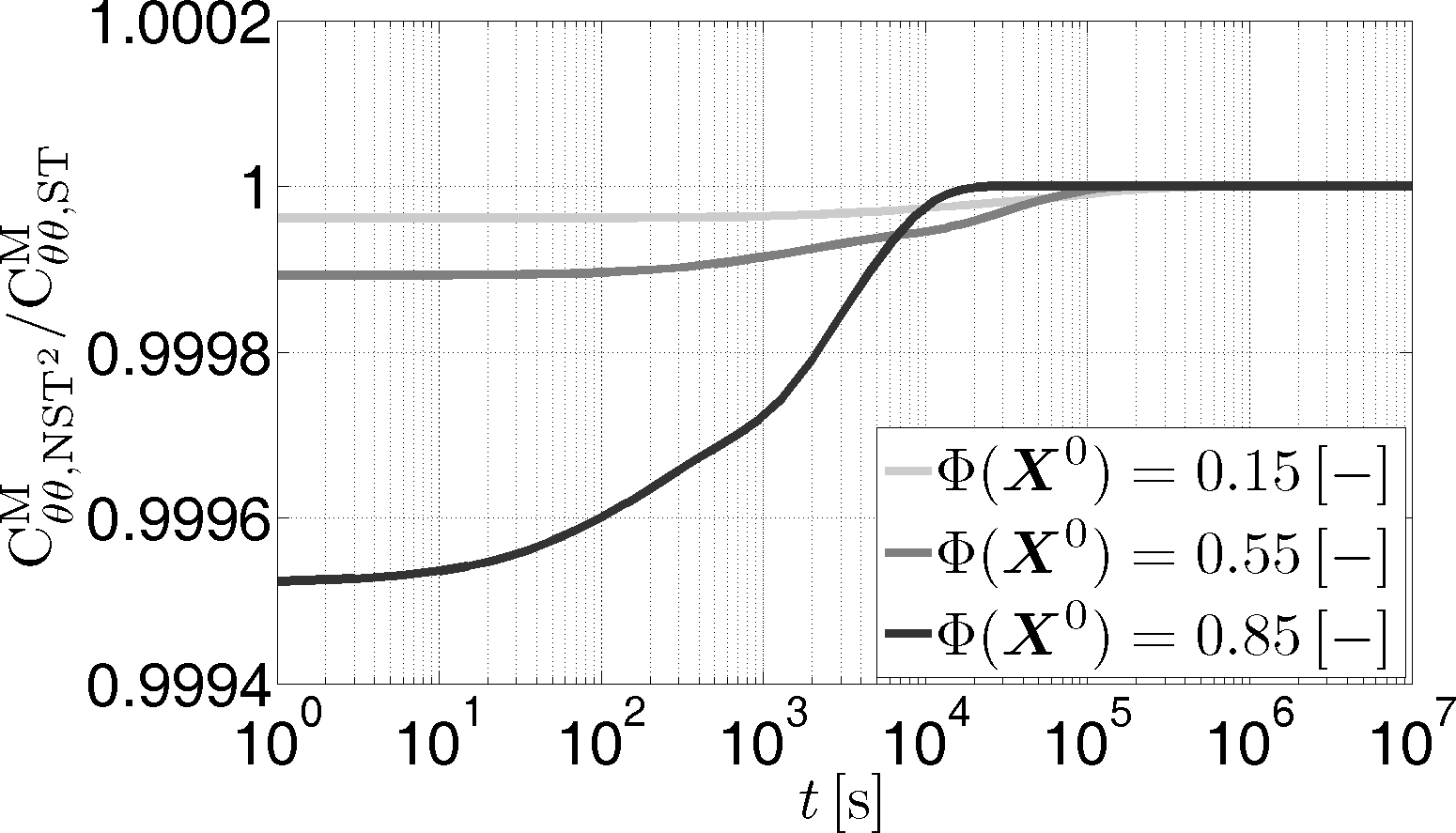}&
\includegraphics*[width=70mm,keepaspectratio]{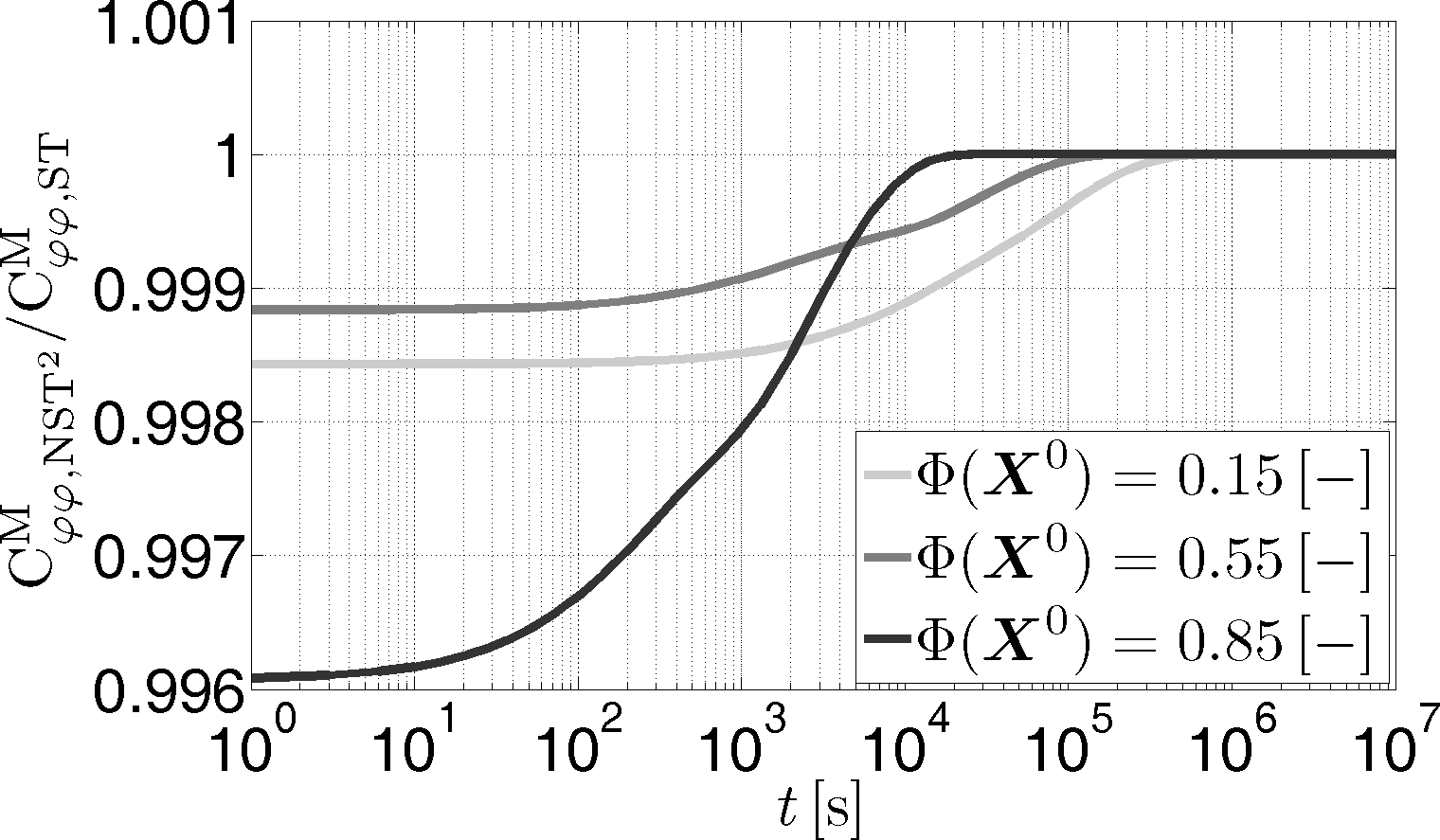}\\
(c)&(d)
\end{tabular}
\end{center}
\caption{Evolution of homogenized material parameters as function of time (PUC):
(a) $\mathrm{K}_{\theta\theta,(1,1)}^{\mathrm{M}}$,
(b) $\mathrm{K}_{\varphi\varphi,(1,1)}^{\mathrm{M}}$,
(c) $\mathrm{C}_{\theta\theta}^{\mathrm{M}}$,
(d) $\mathrm{C}_{\varphi\varphi}^{\mathrm{M}}$
} \label{fig:KC-evol}
\end{figure}
%%%%%%%%%%%%%%%%%%%%%%%%%%%%%%%%%%%%%%%%%%%%%%%%%%%%%%%%%%%%%%%%%

Finally, to address the influence of the degree of material
heterogeneity we repeated the same study employing the irregular
periodic unit cell (SEPUC in
Figure~\ref{fig:scheme}). Figures~\ref{fig:SEPUC-PUC-KMM-CMM-evol}(a)(b)
show evolution of relative humidity along the SEPUC centerline at
various times clearly identifying the material boundaries. While the
evolution trend is similar to the results presented in
Figure~\ref{fig:meso-evol} for PUC, the time to attain steady state
solution slightly increases. Evolutions of the selected effective
moisture terms is plotted in
Figures~\ref{fig:SEPUC-PUC-KMM-CMM-evol}(c)(d) confirming again a
negligible difference between steady state and transient analyses.
While the degree of heterogeneity is pronounced only slightly, the
influence of initial conditions on the prediction of effective
properties is significant.

%%%%%%%%%%%%%%%%%%%%%%%%%%%%%%%%%%%%%%%%%%%%%%%%%%%%%%%%%%%%%%%%%
\begin{figure} [ht!]
\begin{center}
\begin{tabular}{c@{\hspace{5mm}}c}
\includegraphics*[width=70mm,keepaspectratio]{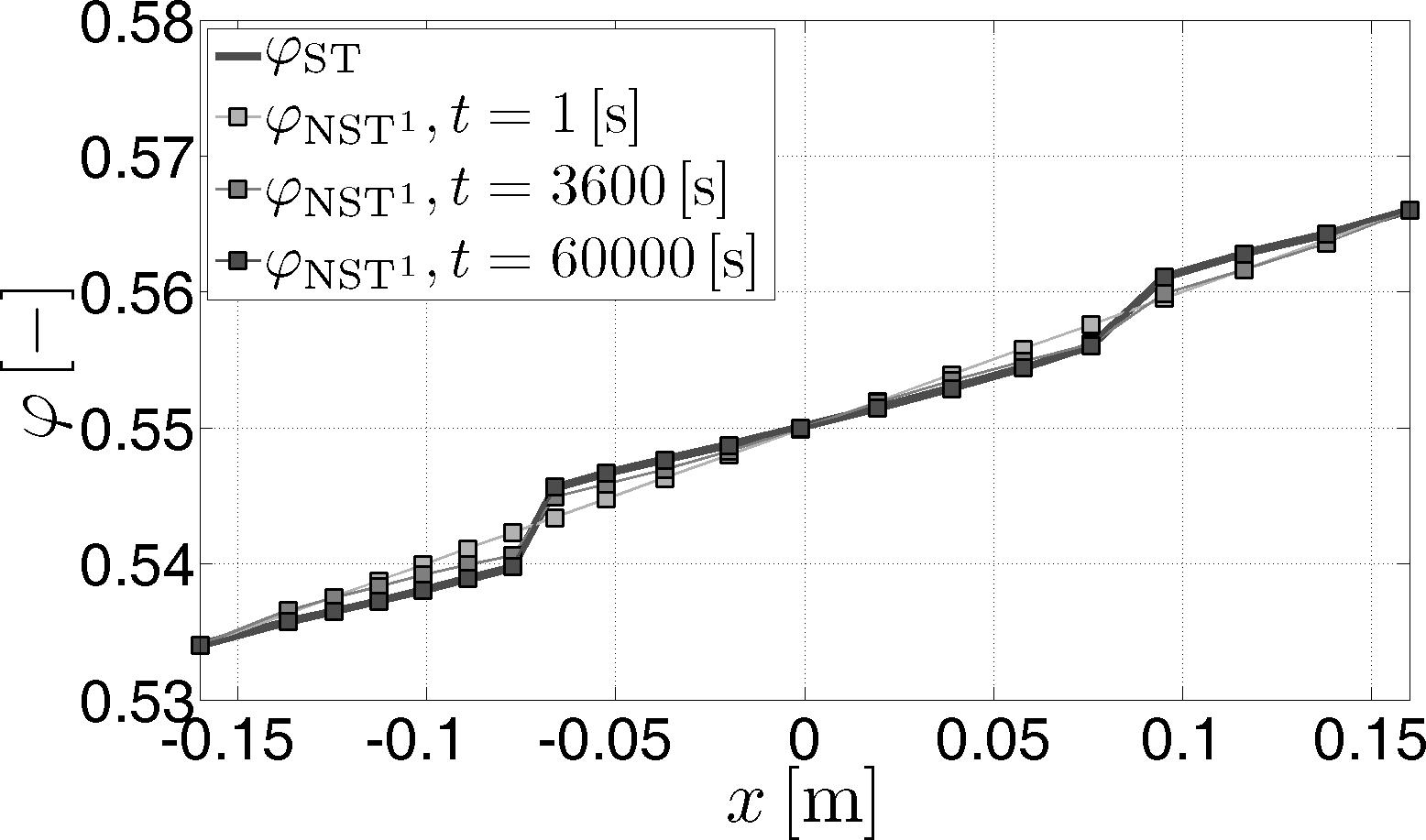}&
\includegraphics*[width=70mm,keepaspectratio]{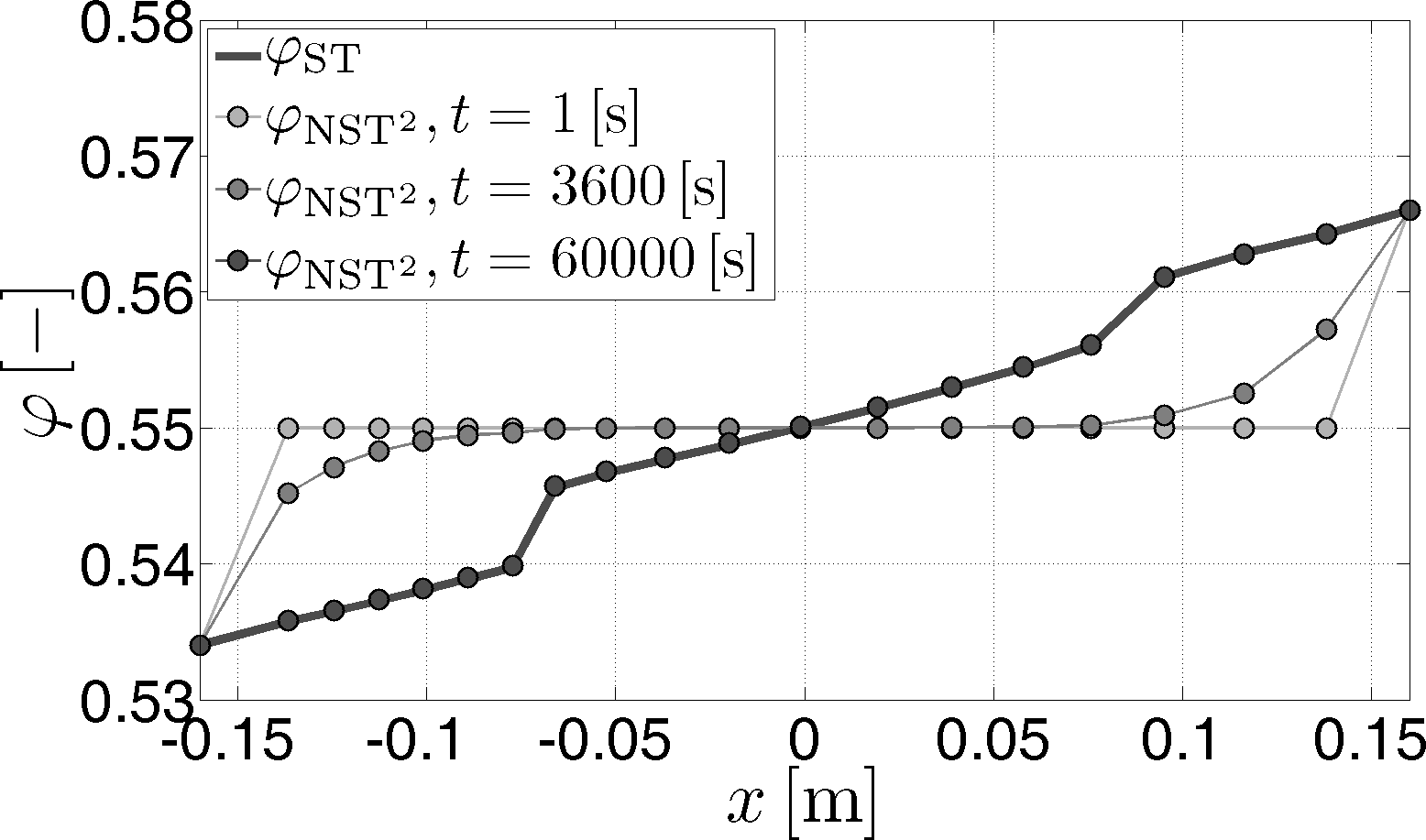}\\
(a)&(b)\\
\includegraphics*[width=70mm,keepaspectratio]{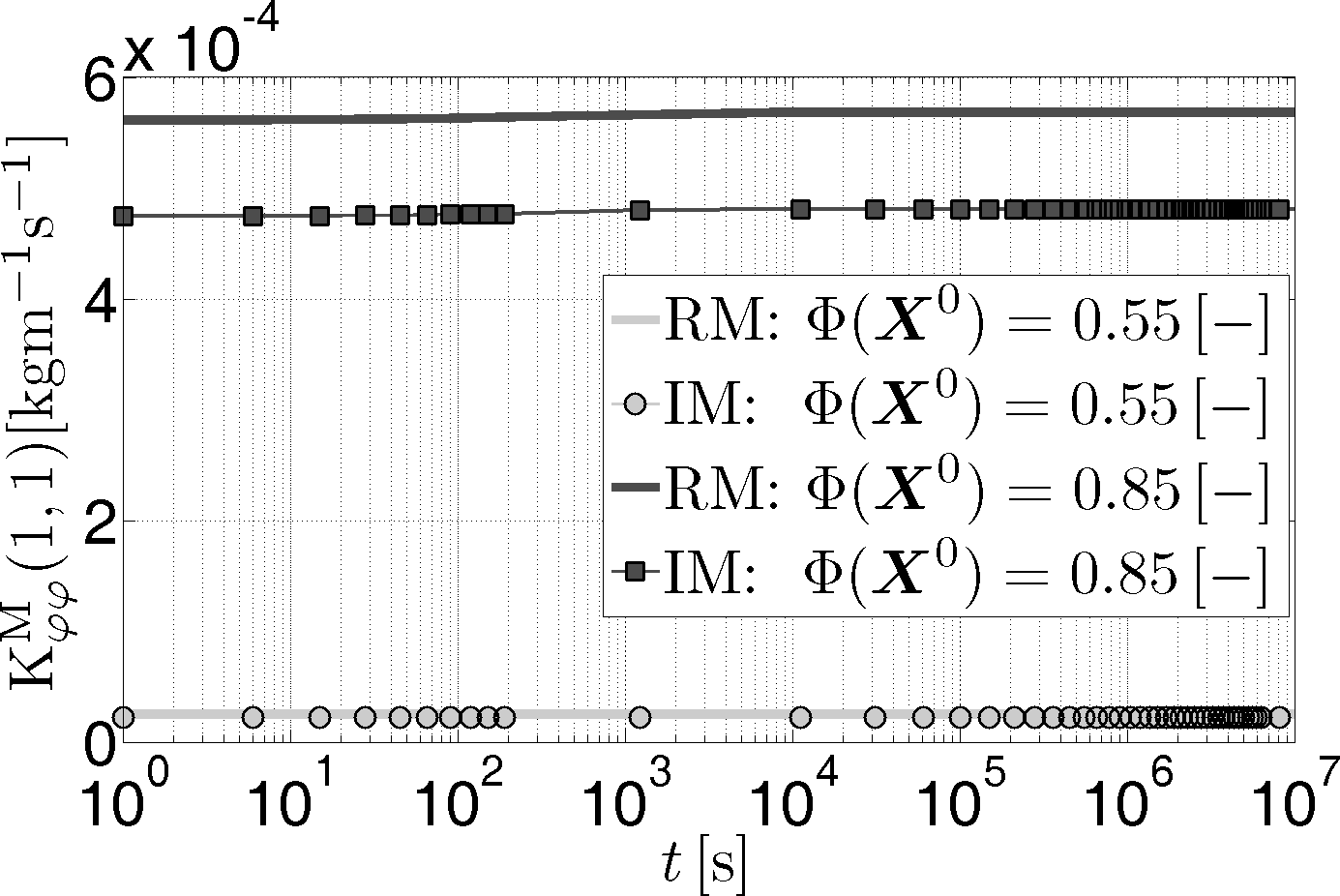}&
\includegraphics*[width=70mm,keepaspectratio]{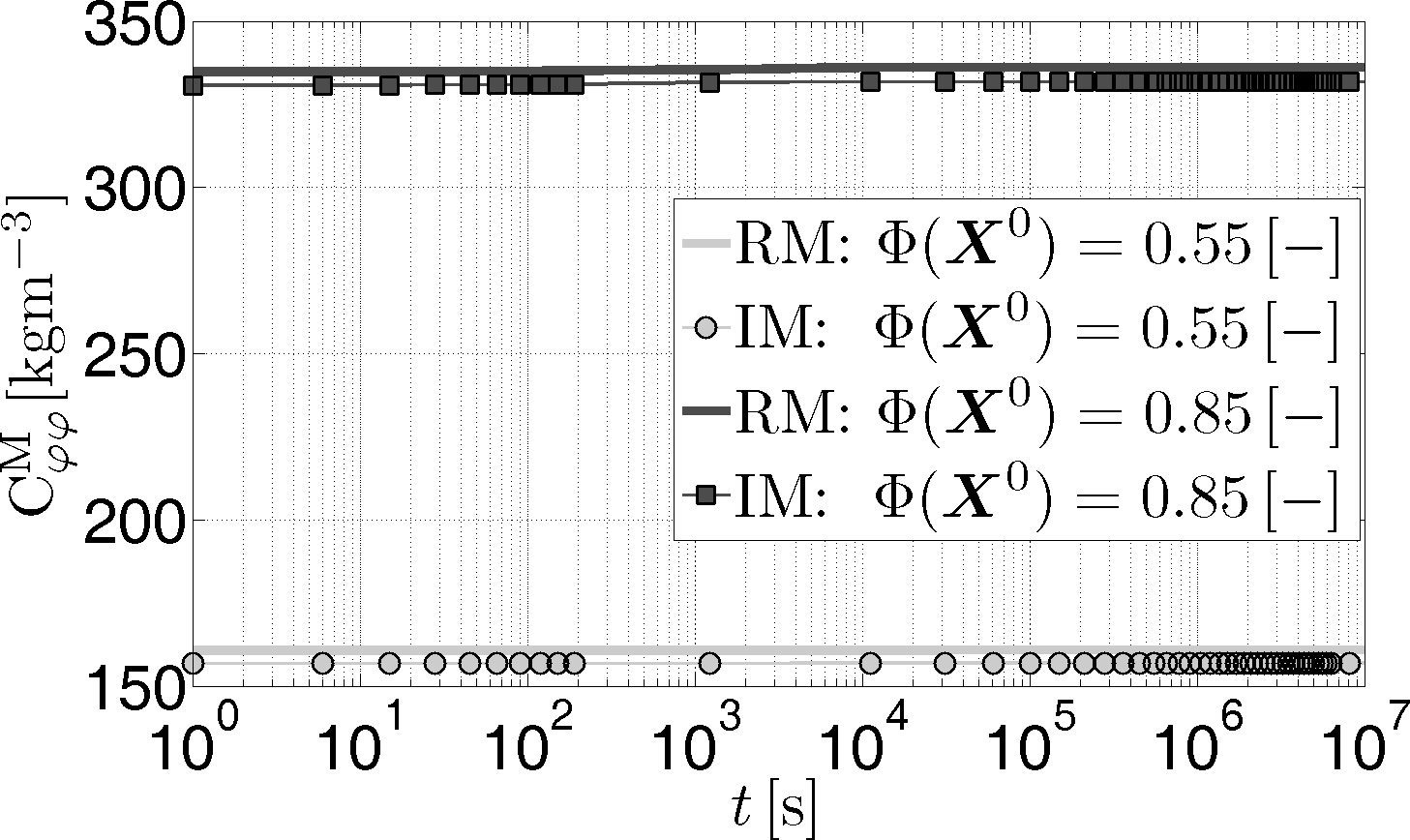}\\
(c)&(d)
\end{tabular}
\end{center}
\caption{Evolution of local moisture (SEPUC): (a) loading conditions I, (b) loading conditions II;
Evolution of homogenized material parameters as s function of time: (c) $\mathrm{K}_{\varphi\varphi,(1,1)}^{\mathrm{M}}$, (d) $\mathrm{C}_{\varphi\varphi}^{\mathrm{M}}$}
\label{fig:SEPUC-PUC-KMM-CMM-evol}
\end{figure}
%%%%%%%%%%%%%%%%%%%%%%%%%%%%%%%%%%%%%%%%%%%%%%%%%%%%%%%%%%%%%%%%%

\subsubsection*{Multi-scale analysis}
%%%%%%%%%%%%%%%%%%%%%%%%%%%%%%%%%%%%%%%%%%%%%%%%%%%%%%%%%%%%%%%%%
Combining all the previous results suggests that for a reasonably
small macroscopic time increment (from one to two hours sufficient to
reach the steady state conditions on the meso-scale) the influence of
non-local terms in Eqs.~\eqref{eq:hom11} and~\eqref{eq:hom12} should
be negligible. Therefore, the macroscopic response should be invariant
with respect to the adopted analysis carried out on meso-scale whether
the transient or steady state. This is evident from the results
plotted in Figure~\ref{fig:evolution} showing evolution of macroscopic
temperatures and relative humidities at selected nodes of mesh, see
Figure~\ref{fig:scheme}.

%%%%%%%%%%%%%%%%%%%%%%%%%%%%%%%%%%%%%%%%%%%%%%%%%%%%%%%%%%%%%%%%%
\begin{figure} [ht!]
\begin{center}
\begin{tabular}{c@{\hspace{5mm}}c}
\includegraphics*[width=70mm,keepaspectratio]{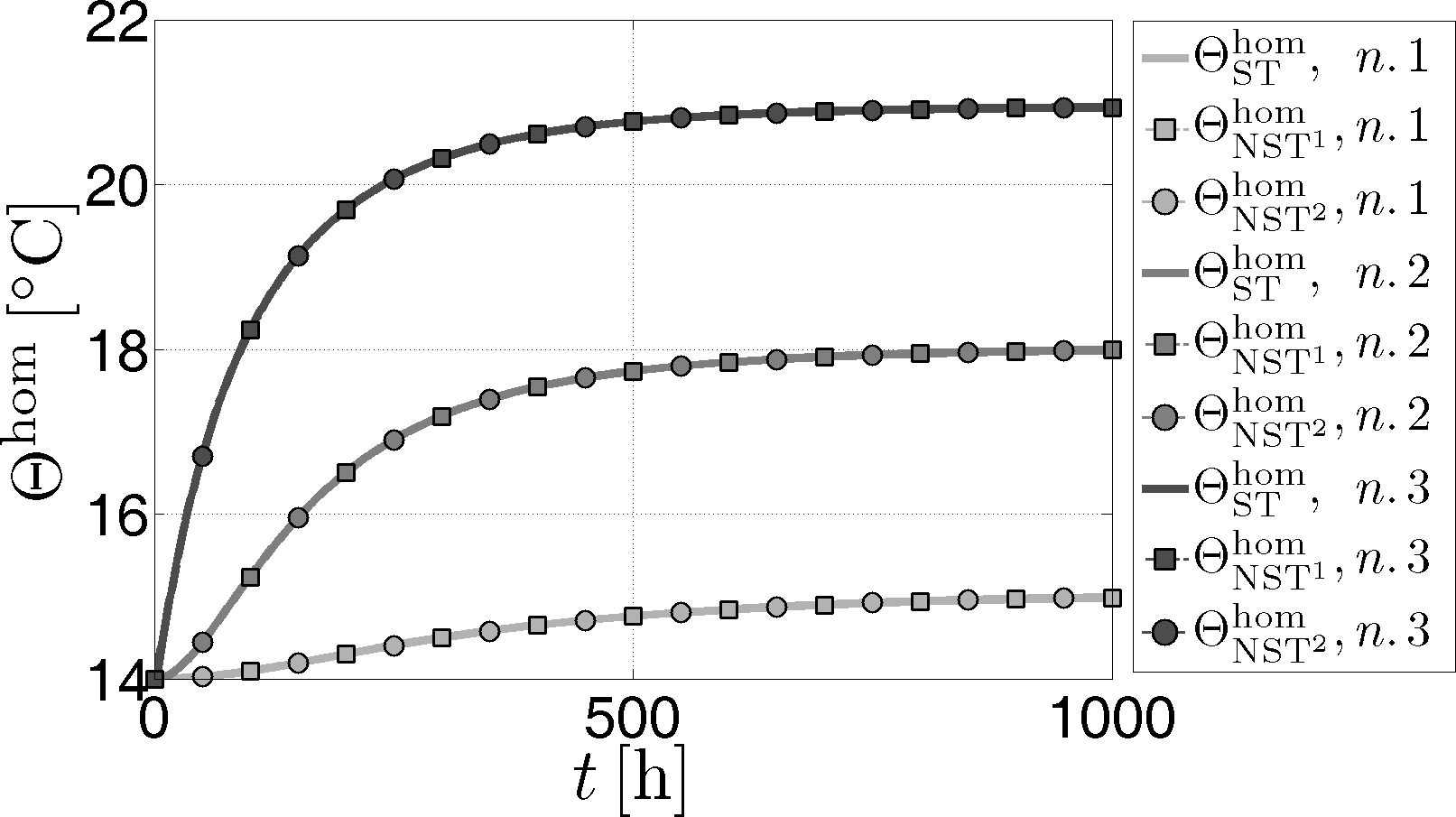}&
\includegraphics*[width=70mm,keepaspectratio]{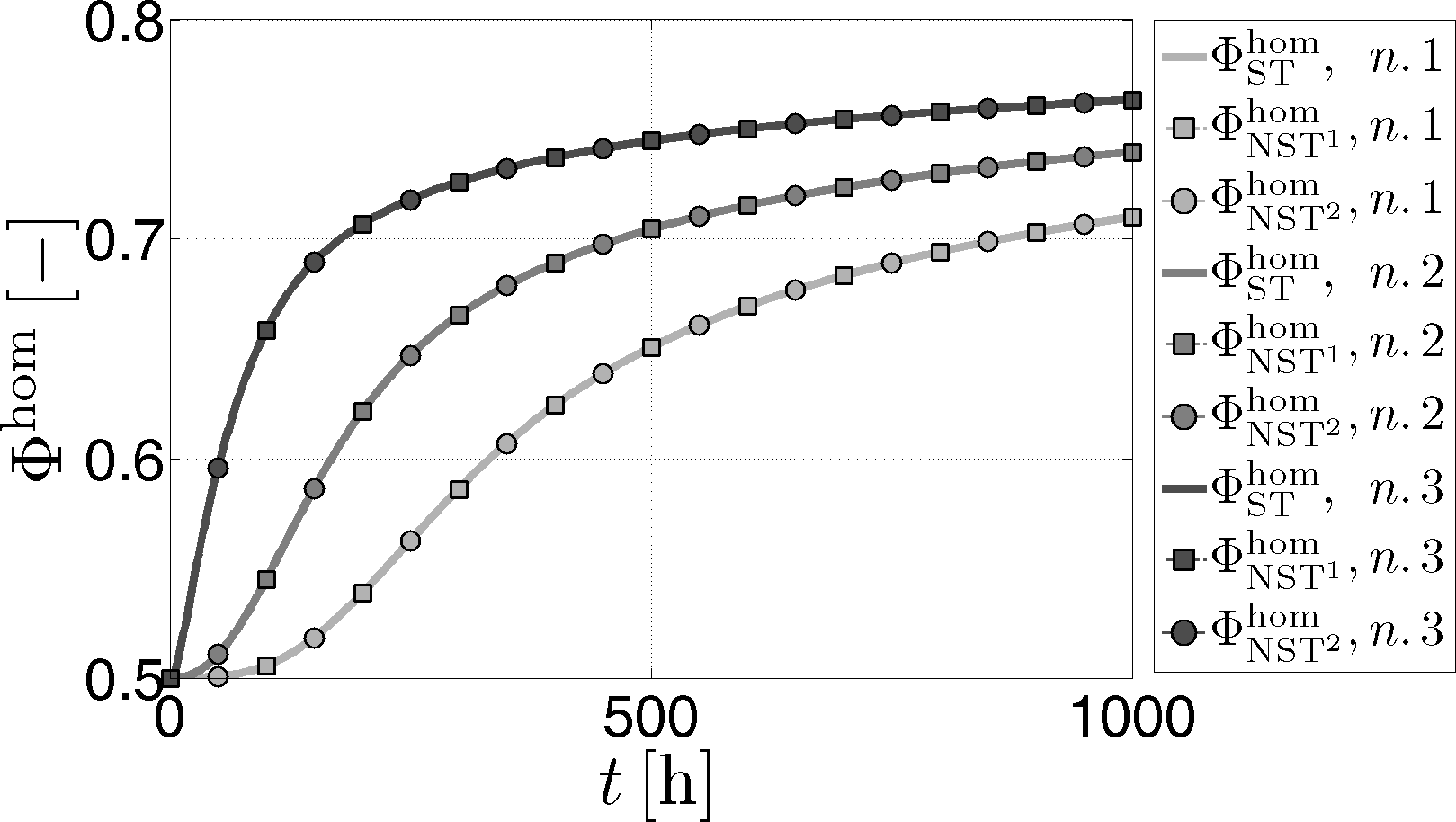}\\
(a)&(b)
\end{tabular}
\end{center}
\caption{Comparison of different macrostructural computations at
selected nodes - (a) temperature evolution, (b) moisture
evolution} \label{fig:evolution}
\end{figure}
%%%%%%%%%%%%%%%%%%%%%%%%%%%%%%%%%%%%%%%%%%%%%%%%%%%%%%%%%%%%%%%%%

This result is quite encouraging particularly with reference to the
analysis of large historical structures, since avoiding a transient
analysis on each mesoscopic unit cell may considerably reduce the
computational cost. 

\section{Conclusions}
%%%%%%%%%%%%%%%%%%%%%%%%%%%%%%%%%%%%%%%%%%%%%%%%%%%%%%%%%%%%%%%%%%%%%%%%%%%%%%%%%%%%%%%%%%%%%%%%%%%%%%%
The present paper gives a brief overview of two particular aspects of
the modeling of historical masonry structures, which show a certain
degree of irregularity on the meso-scale. This observation promotes
application of so called Statistically Equivalent Periodic Unit Cell
being sufficiently simple in comparison to, yet sufficiently
representative of, a real masonry. This issue was addressed first in
\secref{sec:geo_model} to see that under steady state and linear
conditions the degree of heterogeneity, associated with a given
meso-structure, may not play a significant role in the prediction of
effective macroscopic response, recall
\tabref{tab:homog_conductivites}. On the contrary, the morphological
details will become important once considering a nonlinear response
crucially dependent on the actual distribution of local fields.

The latter comment was partially examined next in \secref{sec:coupled}
devoted to the nonlinear fully coupled multi-scale analysis of
simultaneous heat and moisture transport. The principal result of this
study is seen in the possibility of deriving the instantaneous, time
dependent, macroscopic response from a steady-state analysis performed
on the lower scale reflecting all morphological details. However, keep
in mind that this finding is strictly valid for the present problem
being a collection of the selected climatic conditions, material
composition and the assumed constitutive model, and should not be
generalized. For other cases the theoretically predicted dependence of
macroscopic response on the actual RVE size may prove
non-negligible~\cite{Larsson:2010:IJMNE,Sykora:JCAM:2011}.

It is our present interest to exploit the two advancement in
computational efficiency (SEPUC, steady state meso-scale problem) in
the analysis of full scale 3D model of Charles Bridge. Special
attention will be devoted to the implementation efficiency in the
framework of hybrid parallel computing.

%%%%%%%%%%%%%%%%%%%%%%%%%%%%%%%%%%%%%%%%%%%%%%%%%%%%%%%%%%%%%%%%%%%%%%%%%%%%%%%%%%%%%%%%%%%%%%%%%%%%%%%
\section*{Acknowledgment}\noindent
The financial support of the GA\v{C}R grants
P105/11/0411 (J.S. and J.Z.) and P105/11/0224 (M.\v{S}.) is gratefully
acknowledged. In addition, work by J.Z. was partially supported by the
European Regional Development Fund in the IT4Innovations Centre of Excellence
project (CZ.1.05/1.1.00/02.0070). We also thank Jaroslav Vond\v{r}ejc (CTU in
Prague) for his assistance with the FFT-based simulations.

\end{document}

%% file: format.tex
\newcommand{\figref}[1]{\figurename~\ref{#1}}
\newcommand{\secref}[1]{Section~\ref{#1}}
\newcommand{\tabref}[1]{\tablename~\ref{#1}}
\newcommand{\Eref}[1]{Eq.~\eqref{#1}}

\newcommand\de{\,{\mathrm d}}

\newcommand{\bmath}[1]{\mbox{\boldmath$#1$}}
\newcommand{\vek}[1]{\bmath{#1}}

\newcommand{\trn}{{\sf ^T}}

\newcommand{\unitConduct}{\mathrm{Wm}^{-1}\mathrm{K}^{-1}}

\newcommand{\volfrac}{\phi}

\newcommand{\mat}[1]{\mathchoice{\displaystyle\mathbf#1}
{\textstyle\mathbf#1}{\scriptstyle\mathbf#1}
{\scriptscriptstyle\mathbf#1}}

\newcommand{\measure}[1]{|#1|}
\newcommand{\emtrx}[1]{\left[\mathsf{#1}\right]}